# Wide-range continuous tuning of the thermal conductivity of La$_{0.5}$Sr$_{0.5}$CoO$_{3-\delta}$ films *via* room-temperature ion-gel gating

Yingying Zhang[1], William M. Postiglione[2], Rui Xie[3], Chi Zhang[1], Hao Zhou[3], Vipul Chaturvedi[2], Kei Heltemes[2], Hua Zhou[4], Tianli Feng[3], Chris Leighton[2*] and Xiaojia Wang[1*]

[1]*Department of Mechanical Engineering, University of Minnesota, Minneapolis, MN55455, USA*

[2]*Department of Chemical Engineering and Materials Science, University of Minnesota, Minneapolis, MN55455, USA*

[3]*Department of Mechanical Engineering, University of Utah, Salt Lake City, Utah 84112, USA*

[4]*Advanced Photon Source, Argonne National Laboratory, Lemont, Illinois 60439, USA*

**Abstract:** Solid-state control of the thermal conductivity of materials is of exceptional interest for novel devices such as thermal diodes and switches. Here, we demonstrate the ability to *continuously* tune the thermal conductivity of nanoscale films of La$_{0.5}$Sr$_{0.5}$CoO$_{3-\delta}$ (LSCO) by a factor of over 5, *via* a *room-temperature* electrolyte-gate-induced non-volatile topotactic phase transformation from perovskite (with $\delta \approx 0.1$) to an oxygen-vacancy-ordered brownmillerite phase (with $\delta = 0.5$), accompanied by a metal-insulator transition. Combining time-domain thermoreflectance and electronic transport measurements, model analyses based on molecular dynamics and Boltzmann transport, and structural characterization by X-ray diffraction, we uncover and deconvolve the effects of these transitions on heat carriers, including electrons and lattice vibrations. The wide-range continuous tunability of LSCO thermal conductivity enabled by low-voltage (below 4 V) room-temperature electrolyte gating opens the door to non-volatile dynamic control of thermal transport in perovskite-based functional materials, for thermal regulation and management in device applications.

*Authors to whom correspondence should be addressed. Electronic mail: leighton@umn.edu; wang4940@umn.edu

## Introduction

Perovskite oxides with nominal general formula $ABO_3$ are well known for their immensely tunable structures and compositions, and thus physical and chemical properties, making them attractive for applications in superconductivity[1], catalysis[2], various memory devices[3], solid oxide fuel cells[4], *etc*. A and B site cations can be readily substituted to tailor the crystal structure, electronic structure, and doping, while oxygen stoichiometry (*e.g.*, oxygen deficiency $\delta$) adds additional control, particularly *via* oxygen vacancies ($V_O$s). Recently, *active* control of the oxygen stoichiometry in perovskites has come into the spotlight, enabling reversible voltage control of electronic, magnetic, and optical properties[5-9], with much technological potential. Electrochemical operating mechanisms[5], often present in electrolyte-gated perovskite-based devices employing ionic liquid or gel electrolytes, provide one powerful approach to this, where perovskite cobaltites have emerged as prototypical targets. In strontium cobaltite, $SrCoO_{3-\delta}$, for example, a reversible non-volatile topotactic phase transformation between perovskite (P) $SrCoO_{3-\delta}$ (with randomly distributed $V_O$) and oxygen-vacancy-ordered brownmillerite (BM) $SrCoO_{2.5}$ has been achieved through electrolyte gating[10-12]. Due to the highly contrasting electronic and magnetic ground states of these two phases, voltage-driven cycling drives metal-insulator and ferromagnetic-antiferromagnetic transitions, as well as large changes in optical properties, drawing increasing attention[10-13]. Recent work by some of us established that such phenomena are in fact possible throughout the entire $La_{1-x}Sr_xCoO_{3-\delta}$ phase diagram, providing control over the threshold voltage for the P to BM transformation (*via x*), and circumventing issues with the air stability of P-$SrCoO_3$ ($La_{0.5}Sr_{0.5}CoO_3$, for example, is much more air-stable)[14]. In general, the electronic and magnetic properties of $La_{1-x}Sr_xCoO_{3-\delta}$ have been extensively explored and manipulated *via* electrolyte gating, *via* both electrostatic and electrochemical mechanisms[6,14-16].

While the above focuses on structural, electronic, magnetic, and optical properties, electrolyte gating is also attractive for voltage-based control of *thermal* properties, and associated applications. The latter include thermal diodes, thermal switches, and energy conversion and storage[17,18], wherein the thermal conductivity of materials is dynamically tuned to achieve desired performance. For example, Cho *et al.* demonstrated reversible voltage-based tuning of the thermal conductivity of the battery cathode material $Li_xCoO_2$, from 3.7 to 5.4 W m$^{-1}$ K$^{-1}$, *via* electrochemical modulation[19]. More recently, Lu *et al.* voltage tuned the thermal conductivity of $SrCoO_{3-\delta}$ by a factor of ~2.5 by transforming (oxidizing) as-deposited BM-$SrCoO_{2.5}$ to P-$SrCoO_{3-\delta}$, and by a further factor of ~4 by transforming (reducing) BM-$SrCoO_{2.5}$ to a hydrogenated H-$SrCoO_{2.5}$ phase, both *via* non-volatile electrolyte gating[10]. Notably, the factor of 4 tuning ratio of thermal conductivity *via* a bi-state (BM phase → H phase) single-step tuning process appears to be a maximum among literature experimental studies[10,19-28]. While groundbreaking, it should be noted, however, that this work was not able to achieve *continuous* control of $\delta$ between P-$SrCoO_{3-\delta}$ and BM-$SrCoO_{2.5}$ by room-temperature (RT) electrolyte gating, that reversibility and speed remain open questions (particularly when H is involved), and that electronic contributions to the thermal conductivity tuning were found to be negligible, limiting the magnitude of the thermal conductivity tuning ratio[10].

In order to explore the limits of voltage-based tuning of thermal conductivity *via* electrolyte gating between P and BM phases, in this work we focus on $La_{0.5}Sr_{0.5}CoO_{3-\delta}$ (LSCO) as a model system. Critically, this system enables electrolyte gating from an *as-deposited* P phase with pristine metallic electrical conductivity, to an insulating BM phase, maximally modulating the electrical resistivity, with excellent air stability in both phases[14]. Such an approach has *not* been reported with $SrCoO_{3-\delta}$. We combine time-domain thermoreflectance (TDTR) measurements with high-resolution X-ray diffraction (XRD) and electronic transport measurements to explore the thermal conductivity tuning of LSCO through the topotactic P → BM transformation as a function of gate voltage ($V_g$). The $V_g$-driven cubic P to orthorhombic BM transformation and associated metal-insulator transition are found to enable *continuous*, *RT* tuning of the thermal conductivity of LSCO films by a remarkable factor of over 5, a record value for such an approach. Combining temperature (*T*)-dependent thermal measurements with model analyses based on molecular dynamics (MD) simulations and the Boltzmann transport equation (BTE), and comparing to accompanying electrical transport measurements, we further establish the scattering mechanisms of heat carriers that enable this tuning of thermal conductivity. The relative impacts of the Sr substitution, the $V_g$-driven metal-insulator transition (*i.e.*, the electronic contribution to the thermal conductivity), $\delta$ variation, crystal structure modification, and lattice symmetry change, are all discussed in detail.

## Structural characterization of the P → BM phase transformation

In this work, ~45-nm-thick films of P-LSCO epitaxially deposited on LaAlO$_3$(001) (LAO) substrates were driven to the V$_O$-ordered BM structure (BM-LSCO, $\delta \sim 0.5$) by removing oxide ions (O$^{2-}$) using an all-solid-state side-gated electrochemically-operating electrolyte-gate device, shown schematically in Fig. 1a. Application of positive $V_g$ accumulates cations in the ion-gel electrolyte at the interface with the P-LSCO, producing a large electric field and removing O$^{2-}$ from the LSCO film (equivalently, introducing V$_O$s)[14], in a non-volatile fashion. It is known from prior work that this mechanism initially reduces P-LSCO (*i.e.*, increases $\delta$), then leads to the formation of BM-LSCO in a matrix of P-LSCO (*i.e.*, a P-BM mixed phase), and then finally to pure BM-LSCO, as a function of $V_g$[14]. The P and BM structures are shown schematically in Fig. 1b, where the perovskite phase (left) is fully oxygenated (with CoO$_6$ octahedra), while the BM phase (right) consists of perovskite layers separated by comparatively oxygen-deficient $CoO_4$ tetrahedra[10-12,14,29]. The $V_{OS}$ in BM are ordered not only in alternate planes along the *c*-axis, but also in lines in the *a-b*-plane. The P → BM transformation thus quadruples the *c*-axis lattice parameter of the original P-LSCO and lowers the lattice symmetry from cubic to orthorhombic[10-12,14,29].

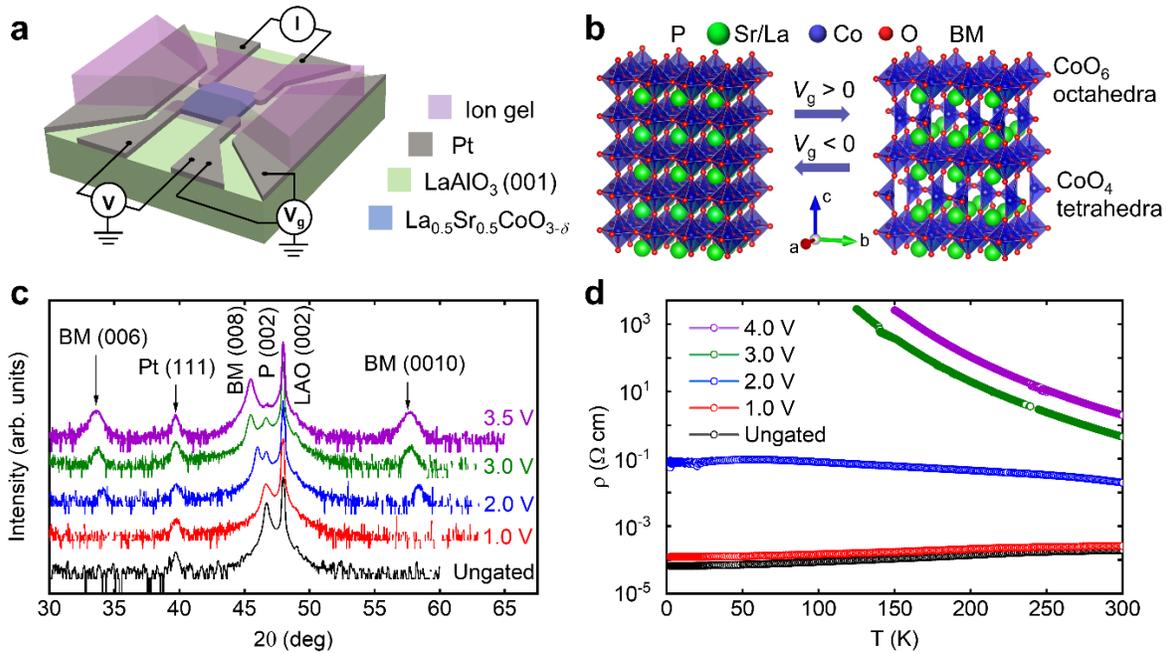

**Fig. 1 | Electrolyte gating and the P → BM phase transformation. a** Schematic of the $La_{0.5}Sr_{0.5}CoO_{3-\delta}$-film-based side-gated electrolyte-gate devices used in this study, with an overlying ion gel as the electrolyte. *I* and *V* are the current and voltage for measurement of four-wire channel resistance, while $V_g$ is the gate voltage. **b** Crystal structure of P-LSCO (left) and BM-LSCO (right). When $V_g$ is applied, the LSCO film is transformed from P → BM and *vice-versa* depending on the $V_g$ polarity: $V_g > 0$ drives to BM, and $V_g < 0$ drives to P. **c** X-ray diffraction from LSCO devices on LAO(001) substrates after gating to various $V_g$. The *y*-axis represents the XRD intensity in arbitrary units, plotted on a logarithmic scale and offset for better visualization. **d** *T*-dependence of the resistivity (from a four-wire measurement) for selected LSCO films after gating to various $V_g$.

Figure 1c shows high-resolution wide-angle specular XRD scans of representative LSCO film devices after ion-gel gating to various $V_g$ from 0 to 3.5 V. Ion-gel gating was performed by sweeping $V_g$ at 0.5 mV s$^{-1}$ to the target voltage. Two-wire transport measurements were made during the sweep, while four-wire $T$-dependent transport and TDTR measurements were made *ex situ*, after bias removal, enabled by the non-volatility of the gate effect[10-12,14,29]. Purely to facilitate TDTR measurements (which require the deposition of an overlying metallic transducer film), each scan here is from a different ion-gel-gated sample. The ungated (black) and +3.5 V (purple) scans were performed on larger-channel-area devices ($4 \times 3.5$ mm$^2$), specifically for better XRD signal-to-noise ratio (SNR), while the other (intermediate-voltage) scans (1.0 V, 2.0 V, and 3.0 V) were performed on $1 \times 1$ mm$^2$ channel devices, as used for four-wire transport measurements. The ungated data (black) reveal the expected phase-pure epitaxial P-LSCO on the LAO(001) substrate (in addition to a Pt(111) peak from the device electrodes), consistent with previous reports on LSCO deposited *via* high-pressure-oxygen sputtering[6,14-16,30]. Additional characterization of P-LSCO films can be found in Supplementary Section 1 and Supplementary Fig. 1, establishing high epitaxial quality and full strain to the substrate.

Moving to the gated-LSCO film XRD patterns in Fig. 1c, we find a clear downshift of the P(002) peaks with increasing $V_g$, indicating expansion of the P-LSCO $c$-axis lattice constant due to $V_O$ formation in the P phase[14,15] (more information on $c$-axis lattice parameters *vs.* $V_g$ can be found in Supplementary Table 1). At 2 V and beyond, a (001)-oriented BM structure is then detected *via* the emergence of additional peaks around 34 and 58°, due to the BM(006) and BM(0010) reflections. This signifies the quadrupling of the unit cell in the $V_O$-ordered BM phase. Initially, these BM-LSCO peaks occur along with P(002) peaks, indicating a mixed phase region, as observed previously and interpreted in terms of phase coexistence across a first-order P → BM transformation[14]. At 3 V, however, the P(002) peak continues to downshift, indicating further lattice expansion, additional $V_O$ formation, and near-complete transformation to BM (the P(002) reflection then becomes the BM(008)). Concomitantly, the BM(006) and (0010) peaks intensify, while the P(002) peak diminishes. By 3 V, the intensity ratio of the BM(008) to BM(006) reflections is in fact ~70, relatively close to the value of ~40 previously observed for fully transformed BM-LSCO after ion-gel gating[14]. Interestingly, little difference is then observed between films gated to 3 and 3.5 V. We interpret this as resistance to further removal of $O^{2-}$ beyond the BM-LSCO stoichiometry (*i.e.*, $\delta \sim 0.5$) at voltages above 3 V. Regarding the structural perfection of the gated BM-LSCO, more detailed analysis (peak fitting) of the 3.5 V sample (magenta) on the larger-channel-area-device revealed additional disorder in the $V_O$ sublattice compared to as-deposited BM films (*e.g.*, $SrCoO_{2.5}$) reported in literature[11,12,31-34] (see Supplementary Section 1). It should also be noted that the P(002) peak intensity, although weak, still remains after 3 V. This relatively low-intensity peak arises from the portion of the P-LSCO film buried beneath the Pt contacts, however, which is ungated. Overall, we thus conclude from Fig. 1c that a reduced P phase exists up to ~1 V, beyond which P + BM phase coexistence persists to ~3 V, followed by near phase-pure BM. These findings are in good general agreement with prior LSCO gating studies[14]. More detailed information on these gated LSCO films, including thicknesses, applied $V_g$, and estimated phases can be found in Supplementary Section 1. As a final comment on these data, we note that, consistent with prior work on gated LSCO[14,15], we find no significant evidence of the hydrogenated phase seen in gating studies of $SrCoO_{3-\delta}$. The transformation in this work is thus simply between P and BM phases; it is therefore bi-state not tri-state switching.

## Electronic transport properties of ion-gel-gated LSCO films

Due to the very different electronic characteristics of the LSCO P and BM phases, major changes in electronic transport accompany the $V_g$-induced P → P + BM → BM transformation deduced above. Figure 1d shows electrical resistivity ($\rho$) *vs.* temperature ($T$) data for an ungated LSCO film, as well as films gated to $1 - 4$ V. The ungated LSCO film is a metallic ferromagnet with a residual resistivity $\rho_0 \approx 70$ µΩ cm, a residual-resistivity-ratio of 2.95, and $T_C \approx 220$ K. This is as expected for relatively thick films of La$_{1-x}$Sr$_x$CoO$_{3-\delta}$ with $x = 0.5$, where metallic conduction and long-range ferromagnetism are observed for $x > 0.17$[35-37]. After gating to 1 V, the metallic conduction in the LSCO film is maintained, albeit with slightly higher $\rho$, reflecting the additional V$_O$ density introduced into the P lattice. As electron donors, V$_O$ compensate holes (the dominant charge carriers in LSCO), thus increasing $\rho$. The relatively small change in $\rho$ for LSCO at 1 V compared to ungated LSCO is consistent with Fig. 1c in that the P phase is retained, but with a higher $\delta$. After applying 2 V, however, the RT $\rho$ increases nearly 100-fold, reflecting not only a substantial increase in V$_O$ concentration (larger $\delta$), but also entry into the P + BM phase coexistence region. The unusual form of $\rho$ *vs. T* for the 2-V-gated sample (note the flattening at low $T$) is also consistent with this; lateral inhomogeneity would be expected in this situation and indeed we detected in-plane anisotropy in the orthogonal electrical resistances in four-wire van der Pauw measurements at low $T$. For the samples gated to 3 and 4 V, where XRD reveals near-phase-pure BM (Fig. 1c), the LSCO films then exhibit completely insulating behavior in Fig. 1d, with $\rho$(RT) of ~0.5 and ~2 Ω cm, respectively. This represents a factor of ~$10^4$ between the RT $\rho$ of ungated P-LSCO and BM-LSCO, comparable to literature values for electrolyte-gated SrCoO$_{3-\delta}$ and La$_{0.5}$Sr$_{0.5}$CoO$_{3-\delta}$[12,14].

In addition to these *ex-situ* measurements of $T$-dependent four-wire resistivity (Fig. 1d), *in-situ* measurements of two-terminal resistance (*i.e.*, source-drain resistance) were also made while sweeping to the target $V_g$ for each sample (see Supplementary Fig. 2). These two-terminal measurements show reassuring overall agreement with the *ex-situ* trends in resistivity depicted in Fig. 1d, *i.e.*, minor increases in resistance before ~2 V, a sharp increase between ~2 – 3 V, then a region with relatively weak $V_g$ dependence, indicating completion of the P → P + BM → BM transformation. These data (Supplementary Fig. 2) also highlight the high reproducibility of the gating process, underscoring the robustness of our approach.

## Tuning of LSCO thermal conductivity with gate voltage

The thermal conductivities of LSCO thin films gated to various $V_g$ (from 0 to 4 V) were measured with the TDTR approach, which is an ultrafast-laser-based pump-probe technique[38,39]. All TDTR measurements were made *ex-situ*, after biasing and ion-gel removal, taking advantage of the non-volatile nature of the electrolyte gating in this case[10-12,14,29]. Figure 2a illustrates the sample configuration for TDTR measurements, which consists of a thin layer of Pt deposited on top of the LSCO films, serving as the heat source and thermometer. Pt was chosen as the transducer for this study, instead of the typical Al, specifically to avoid Al-gettering-induced $V_O$ formation in the underlying LSCO[40]. By varying the modulation frequency in the TDTR approach, the measurement sensitivity to the LSCO thermal conductivity and the thermal interfaces can be tailored. This enables us to also extract the interfacial thermal conductance between the Pt transducer film and the LSCO film ($G_1$), and between the LSCO film and the LAO substrate ($G_2$), by combining data reduction of routine TDTR measurements with dual-frequency TDTR analyses (see Supplementary Section 3 for details).

As part of the input parameter set used in the analysis of TDTR data to determine the thermal conductivity of LSCO films, the thickness ($d_{Pt}$) and thermal conductivity ($\Lambda_{Pt}$) of the Pt transducer were obtained, respectively, from grazing incidence X-ray reflectometry and four-point probe measurements of a standard Si/SiO$_2$(300 nm)/Pt(~70 nm) reference sample at RT. For TDTR data reduction as a function of $T$, $\Lambda_{Pt}$ is set based on literature data[41], as is the $T$-dependent volumetric heat capacity of the transducer ($C_{Pt}$)[42]. In addition to the Pt transducer, some dimensional parameters and thermal properties of the LSCO film ($d_{LSCO}$, $C_{LSCO}$) and LAO substrate ($\Lambda_{LAO}$, $C_{LAO}$) are also needed. The $T$-dependent $\Lambda_{LAO}$ data were obtained from TDTR measurements of a bare LAO substrate. $C_{LSCO}$ and $C_{LAO}$ were calculated as a function of $T$ based on the Debye model using the literature reported Debye temperature[43,44], and are in good agreement with literature data[43,45-48] (see details in Supplementary Section 6 and Supplementary Fig. 7). For TDTR measurements at low $T$ ($T < 273$ K) where the material's thermal properties (*i.e.*, specific heat and thermal conductivity) exhibit stronger $T$ dependence, we also performed $T$-dependent corrections, taking into account the impact of steady-state heating and the pump per-pulse heating (details are presented in Methods and Supplementary Section 6).

Figure 2b shows representative TDTR ratio data of the in-phase and out-of-phase thermoreflectance signals ($-V_{in}/V_{out}$) measured on the ungated LSCO film at RT. Three modulation frequencies (1.5, 9.0, and 18.8 MHz) were used to enable tailoring of the measurement sensitivity. The solid lines in Fig. 2b are best fits to the TDTR data based on a heat diffusion model for a multi-layer structure[38,49]. At RT, the fitted through-plane thermal conductivity of the ungated LSCO sample is $4.6 \pm 2.0$ W m$^{-1}$ K$^{-1}$. The interfacial thermal conductances of the Pt/LSCO ($G_1$) and LSCO/LAO ($G_2$) interfaces are determined as 400 and 800 MW m$^{-2}$ K$^{-1}$, respectively. More details about the TDTR thermal measurements and the extraction of $G_1$ and $G_2$ are provided in Methods, and Supplementary Section 3.

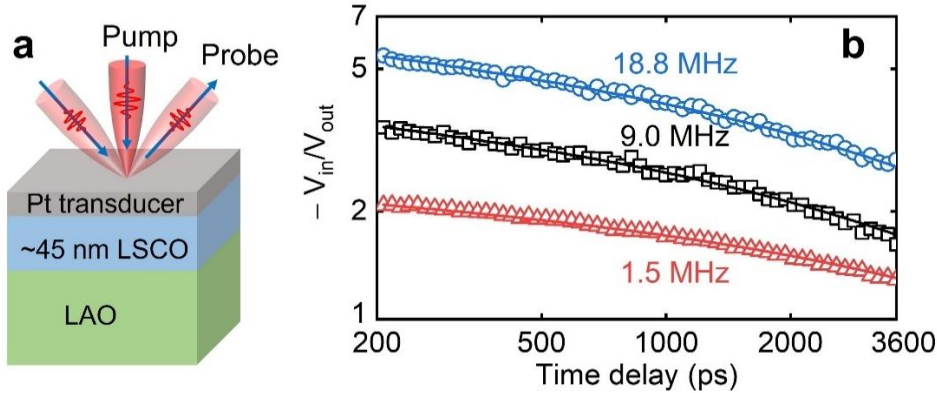

**Fig. 2 | Thermal measurements using the TDTR approach. a** Sample configuration for TDTR measurements. **b** Representative TDTR ratio signals of the ungated LSCO film measured at three modulation frequencies. The solid lines are the best fits to the through-plane thermal measurement data based on a heat diffusion model. The data reduction gives a through-plane thermal conductivity of $4.6 \pm 2.0$ W m$^{-1}$ K$^{-1}$ for this ungated LSCO film.

The RT electrical resistivity and thermal conductivity of LSCO thin films are summarized as functions of $V_g$ in Fig. 3a. As $V_g$ increases from 0 to 4 V, the change in $\rho$ reaches almost five orders of magnitude, a clear reflection of the metal-insulator transition shown in Fig. 1d. Coincident with this, the central experimental result of this work is that the through-plane thermal conductivity of LSCO decreases from $4.6 \pm 2.0$ to $0.85 \pm 0.32$ W m$^{-1}$ K$^{-1}$ as $V_g$ varies from 0 to 4 V, *i.e.*, a factor of more than 5 reduction *via* ion-gel gating. We consider this record-high tuning factor to be for the through-plane thermal conductivity due to possible thermal transport anisotropy in the BM-LSCO film. However, it is reasonable to convert the in-plane electrical resistivity to the through-plane electronic thermal conductivity (see Supplementary Section 7 for more details). We emphasize that the thermal conductivity can be *continuously* tuned over this range, in a *non-volatile* fashion, at *room temperature*. It should be noted that the thermal conductivity of our as-deposited P-LSCO films (with $x = 0.5$ and $\delta \approx 0.1$, see Methods and the discussion below regarding the determination of $\delta$) is $4.6 \pm 2.0$ W m$^{-1}$ K$^{-1}$, which is less than half the equivalent values for several unsubstituted $ABO_3$ perovskites (*e.g.*, 11 W m$^{-1}$ K$^{-1}$ for SrTiO$_3$[50], 13 W m$^{-1}$ K$^{-1}$ for BaSnO$_3$[51], and 13 W m$^{-1}$ K$^{-1}$ for LaAlO$_3$[52]). Also, prior experimental studies reported a bulk thermal conductivity of 6 W m$^{-1}$ K$^{-1}$ for single-crystal La$_{0.7}$Sr$_{0.3}$CoO$_3$[52,53]. The significantly smaller thermal conductivities of La$_{0.7}$Sr$_{0.3}$CoO$_3$ and La$_{0.5}$Sr$_{0.5}$CoO$_3$ are partially attributed to enhanced phonon-defect scattering resulting from the mass mismatch and local strains induced by Sr substitution for La[51-53]. Below, we analyze the change of thermal conductivity versus $V_g$ in detail, including discussing the contributions to the thermal conductivity from various critical factors.

It is important to note, however, that the $V_g$-driven P → BM and metal-insulator transitions also impact thermal transport across the Pt/LSCO interfaces. Figure 3b summarizes the results on $G_1$ extracted from TDTR, where a smaller $G_1$ is observed when LSCO is gated to higher $V_g$. For BM-LSCO, with insulating behavior, heat can be transferred across the Pt/BM-LSCO interface *via* two channels: electron-phonon interactions in Pt and phonon-phonon interactions (between Pt and BM-LSCO phonons). The average measured $G_1$ (170 MW m$^{-2}$ K$^{-1}$) of our BM-LSCO is comparable to the values of interfacial thermal conductance between Pt and oxides reported in literature[54-57] (Supplementary Fig. 4). However, for $V_g < 3$ V, the P phase in the LSCO films provides an additional channel for interfacial heat transfer, *via* electron-electron interactions (between Pt and P-LSCO) across the interface[58,59], thus leading to larger $G_1$. This increase in $G_1$ reduces the measurement sensitivity to $G_1$ (Supplementary Fig. 3). As a result, we used a nominal value of $G_1$ (400 MW m$^{-2}$ K$^{-1}$, black dashed line in Fig. 3b) for LSCO films at $V_g < 2$ V, when measurement sensitivity is insufficient to determine $G_1$ (*i.e.*, the choice of $G_1$ value does not change the fitting results of the LSCO thermal conductivity). Besides, $G_2$ of the LSCO/LAO epitaxial interface is measured to be 800 WM m$^{-2}$ K$^{-1}$ using the dual-frequency TDTR approach with enhanced measurement sensitivity to $G_2$[60] (see Supplementary Section 3 for more details). This value of $G_2$ is consistent with literature data for strongly bonded interfaces[61].

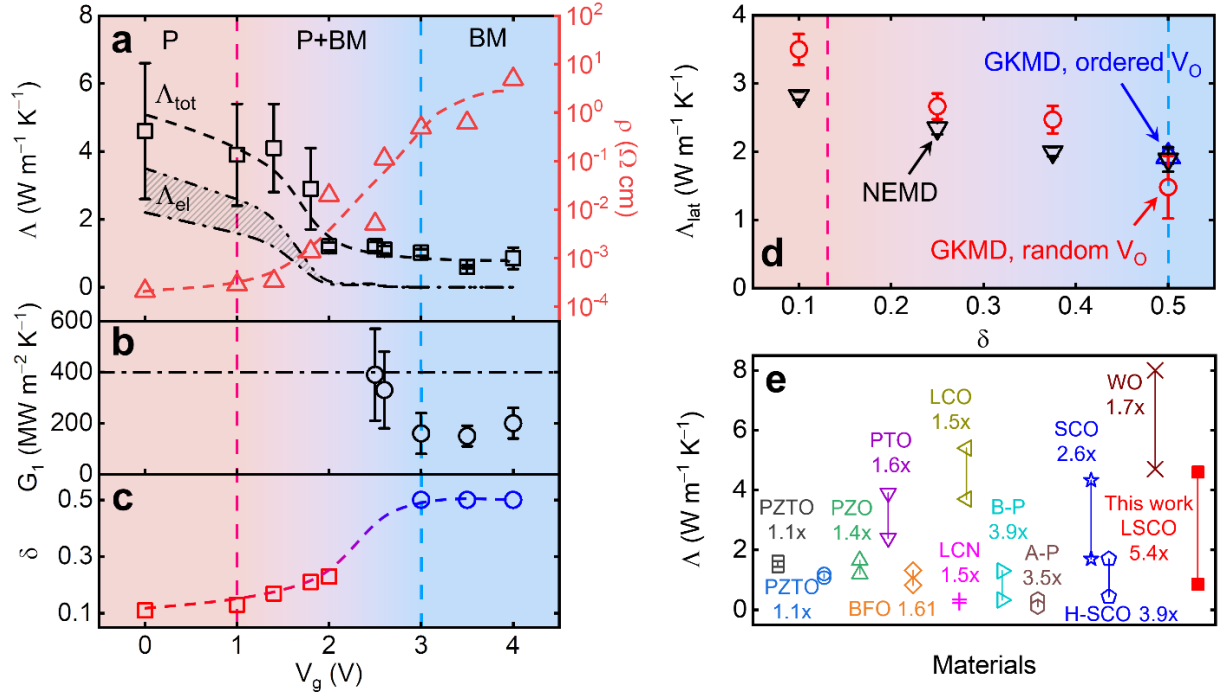

**Fig. 3 | Impact of $V_g$ on thermal properties. a** The thermal conductivity (left axis) and electrical resistivity (right axis) of LSCO films, **b** the interfacial thermal conductance between Pt and LSCO, and **c** oxygen non-stoichiometry due to vacancies, as functions of $V_g$. In panel **a**, $\Lambda_{tot}$ represents the measured total thermal conductivity of the LSCO films. $\Lambda_{el}$ represents the electronic thermal conductivity estimated from electrical conductivity based on the Wiedemann-Franz law. The black dash-dotted lines represent the upper and lower limits of $\Lambda_{el}$ based on different Lorenz numbers ($L$). The dashed lines serve as guides to the eye. In panel **c**, $\delta = 0.5$ (blue circles) is assumed for the complete transition to brownmillerite ($V_g \geq 3$ V), as our method used to estimate $\delta$ (red squares) is not valid above $\delta \approx 0.25$ (see Methods). The red-blue graded dashed line serves as a guide to the eye. **d** MD simulation results, obtained from both GKMD and NEMD, for the lattice thermal conductivity of LSCO with different $\delta$. In all panels, the vertical pink and blue dashed lines represent, respectively, the starting and ending points of the P → P + BM → BM transformations. **e** Comparison of thermal conductivity tuning factor through bi-state tuning process in this work and previous experimental studies[10,19-28], including PbZr$_{0.3}$Ti$_{0.7}$O$_3$ (PZTO, 5 cycles)[20,21], PbZrO$_3$ (PZO, 10+ cycles)[22], PbTiO$_3$ (PTO)[23], BiFeO$_3$ (BFO)[24], bio-polymers (B-P, 10+ cycles)[25], azobenzene polymers (A-P, 6 cycles)[26], liquid crystal networks (LCN)[27], Li$_x$CoO$_2$ (LCO, 2 cycles)[19], WO$_3$ (WO, 3 cycles)[28], and SrCoO$_{3-\delta}$ (SCO, 1 cycle, and H-SCO)[10]. Here, one cycle is defined as the transformation from state/phase A to state/phase B, and then return to state/phase A. Cycle numbers are provided when available from the literature.

## Effects of the P → BM transformation and metal-insulator transition on the thermal conductivity of LSCO

As already noted, the $V_g$-driven P → BM phase transformation in electrolyte-gated LSCO involves several factors, including a large change in $\delta$, lowered crystal symmetry, $V_O$ ordering, and an accompanying metal-insulator transition. The P → BM (cubic to orthorhombic) crystal structure change (Fig. 1b,c), induced by the $V_g$-driven increase in $\delta$, impacts phonon transport, thus modifying thermal conductivity *via* the phonon contribution. To better quantify the impact of $V_O$ on the thermal conductivity in Fig. 3a, it is useful to correlate $V_g$ with $V_O$ concentration. To this end, we estimated $\delta$ using an established empirical approach (see Methods), based on electrical transport data[14,16,52]. The resulting values are plotted in Fig. 3c. The expected increasing trend in $\delta$ is observed for $V_g$ from 0 to 2 V. For films gated above 2 V, however, our approach is no longer applicable (see Methods), and we simply assign the nominal $\delta = 0.5$ implied by our XRD observation of near-phase-pure BM-LSCO at 3, 3.5, and 4 V (blue circles in Fig. 3c).

To gain further insight into the thermal conductivity change in Fig. 3a, we decompose it into electronic and lattice contributions, *i.e.*, $\Lambda_{tot} = \Lambda_{el} + \Lambda_{lat}$ (with $\Lambda_{el}$ being the electronic thermal conductivity and $\Lambda_{lat}$ the lattice thermal conductivity). $\Lambda_{el}$ can be estimated from the electrical conductivity ($\sigma = 1/\rho$) *via* the Wiedemann-Franz law (WFL), *i.e.*, $\Lambda_{el} = L\sigma T$, with $L$ being the Lorenz number. Since the value of $L$ varies non-trivially depending on the nature of the material system, we used the lower and upper limits on $L$ commonly accepted in the literature (1.5 and $2.44 \times 10^{-8}$ $V^2$ $K^{-2}$), to estimate $\Lambda_{el}$[62-65]. The results are plotted as the shaded envelope between black dash-dotted lines in Fig. 3a, exhibiting the expected generally decreasing trend with $V_g$ due to the metal-insulator transition. For the as-deposited P-LSCO film with $\delta \approx 0.1$ (see Fig. 3c), the lower and upper limits of $\Lambda_{el}$ are estimated as 2.2 and 3.5 W $m^{-1}$ $K^{-1}$, resulting in the upper and lower limits of $\Lambda_{lat} = \Lambda_{tot} - \Lambda_{el}$ of 2.4 and 1.1 W m$^{-1}$ K$^{-1}$, respectively. The upper limit of $\Lambda_{lat}$ decreases from 2.4 to 0.85 W m$^{-1}$ K$^{-1}$, a 65% reduction, as $V_g$ increases from 0 to 4 V, at which point the P $\rightarrow$ BM transformation is complete. Importantly, we thus find that electrons contribute comparably with phonons for P-LSCO, which is reasonable for "dirty", *i.e.*, higher resistivity, metals[66]. This is notably different from the P-SrCoO$_{3-\delta}$ reported by Lu *et al.*[10], where the electronic contribution to thermal conductivity was deduced to be negligible. We attribute this difference to the likely larger $\delta$ and disorder level in the P phase in Ref. [10], where the P-SrCoO$_{3-\delta}$ was achieved from as-deposited BM films *via* gating, in contrast to our as-deposited high-epitaxial-quality P-LSCO. Supporting this, Lu *et al.* reported $\rho$(RT) $\approx 5 \times 10^{-2}$ $\Omega$ cm in their P-SrCoO$_3$[10], compared to $\rho$(RT) $\approx 2 \times 10^{-4}$ $\Omega$ cm in our as-deposited P-LSCO, *i.e.*, our P conductivity is ~250 times higher.

To further understand the trend of $\Lambda_{lat}$ versus $V_g$ (and therefore also $\delta$), we performed Green-Kubo MD (GKMD) and nonequilibrium MD (NEMD) simulations of thermal conductivity (see Supplementary Section 5 for calculation details), with results shown in Fig. 3d. The GKMD and NEMD results agree reasonably well. The MD simulations predict a ~62% $\Lambda_{lat}$ reduction (from 3.9 to 1.5 W m$^{-1}$ K$^{-1}$) as $\delta$ increases from 0.1 to 0.5, assuming *random* spatial distributions of V$_{OS}$. Considering that the V$_{OS}$ in the BM phase of our LSCO films are actually ordered,[52] we also calculated the thermal conductivity of orthorhombic BM-LSCO, with ordered V$_{OS}$, using MD simulations; this yields a larger $\Lambda_{lat} = 1.9$ W m$^{-1}$ K$^{-1}$ at $\delta = 0.5$ due to the decrease of phonon-defect scattering. There is thus a ~50% lattice thermal conductivity reduction from $\delta = 0.1$ in the disordered P phase to $\delta = 0.5$ in the ordered BM phase. Importantly, these 62% and 50% reductions in $\Lambda_{lat}$ agree reasonably with the upper limit of 65% obtained from our $\Lambda_{lat} = \Lambda_{tot} - \Lambda_{el}$ extraction. Note that the absolute values of $\Lambda_{lat}$ from the experiment and MD simulations are not directly compared because the classical interatomic potential used in our MD simulations provides only qualitative insights. Based on the above discussion, we conclude that the decrease in thermal conductivity with increasing $V_g$ originates from comparable reductions in both the electronic and lattice contributions. This plays an important role in our record-high (> 5-fold) tuning of thermal conductivity achieved here *via* bi-state gating. Figure 3e summarizes the tuning factor of material thermal conductivities in our work, and in prior experimental studies, *via* a bi-state tuning process[10,19-27]. Evidently, the tuning factor of LSCO thermal conductivity in our work is the largest among experimental reports.

Considering the reversible nature of the $V_g$-induced P $\leftrightarrow$ BM phase transformation in LSCO, we also reverted some gated BM-LSCO films to the P phase by applying reverse gate voltages of up to –4.5 V at RT. While this typically generates an approximately $10^4$ decrease in RT $\rho$ due to the BM $\rightarrow$ P insulator-metal transition, the resulting RT $\rho$ values of recovered P-LSCO films are typically double that of the initial as-deposited P films, corresponding to approximately half the initial electronic thermal conductivity. The thermal measurement of one such film generated a recovered P-LSCO thermal conductivity of 3.5 ± 1.1 W m$^{-1}$ K$^{-1}$ at RT. This ~20% reduction in the $\Lambda_{tot}$ of recovered P-LSCO compared to the as-deposited P-LSCO film no doubt arises from additional structural disorder induced during the cyclic gating (see Supplementary Section 8 for more discussion). Nevertheless, such data confirm the overall reversibility of the approach presented here. Further improvement of reversibility could be achieved through better device design and optimized gating conditions.

## Temperature-dependent thermal conductivities of LSCO

To further clarify the mechanisms of the scattering processes of the relevant heat carriers in LSCO, we also performed *T*-dependent thermal measurements on P-LSCO (ungated) and BM-LSCO ($V_g$ = 3 V) from ~90 to ~500 K. Generally, as $T$ increases, the thermal conductivity of a dielectric crystal should first increase (due to the increase of the specific heat)[67] and then decreases with a $1/T$ trend (due to the increase of Umklapp scattering)[68]. However, neither of our LSCO samples exhibit such typical features. Instead, both P-LSCO and BM-LSCO show a monotonically increasing trend with $T$ in our measurement range (Fig. 4a). We interpret this observation in terms of the two relevant types of heat carriers (electrons and phonons), with differing $T$ dependencies. In P-LSCO, clearly from the above, both electrons and lattice vibrations contribute to thermal transport. The $\Lambda_{el}$ estimated from the WFL with the lower and upper limits of $L$ (using our measured $\rho(T)$) is presented as the envelope shaded with blue dashed lines in Fig. 4b. It can be seen that $\Lambda_{el}$ increases with $T$, dominating the increase in $\Lambda_{tot}$. It is worth noting that a dip is observed in $\Lambda_{el}$ at ~235 K, however. This is because the Curie temperature of our P-LSCO films is ~230 K[14,37], at which the transition from ferromagnetic to paramagnetic behavior induces a change in the $\rho(T)$ behavior, as can be seen in Fig. 1d upon close inspection.

The $T$-dependent $\Lambda_{lat}$ of P-LSCO was also calculated using the BTE to take into account the size effect in the thin film, and is shown as the green dash-dotted line in Fig. 4b[52]. $\Lambda_{lat}$ exhibits a peak at around 200 K. Generally, the peak temperature of thermal conductivity of insulating single-crystal perovskite oxides in unsubstituted $ABO_3$ perovskites appears around or below 100 K (including $LaAlO_3$, the substrate used in this work, see Supplemental section 6)[43,45-48]. The difference in peak temperature between our P-LSCO and other such $ABO_3$ perovskites arises because of the presence of a significant $V_O$ concentration, the disorder on the A-site (La/Sr), and the sample boundaries, inducing strong $T$-independent extrinsic phonon scattering, leading to a higher peak temperature in $\Lambda_{lat}$ compared to pristine $ABO_3$ crystals. Finally, we summed $\Lambda_{el}$ and $\Lambda_{lat}$ ($\Lambda_{el} + \Lambda_{lat}$, shaded area with red solid lines in Fig. 4b) to compare with $\Lambda_{tot}$ obtained from measurements, obtaining a reasonably good agreement.

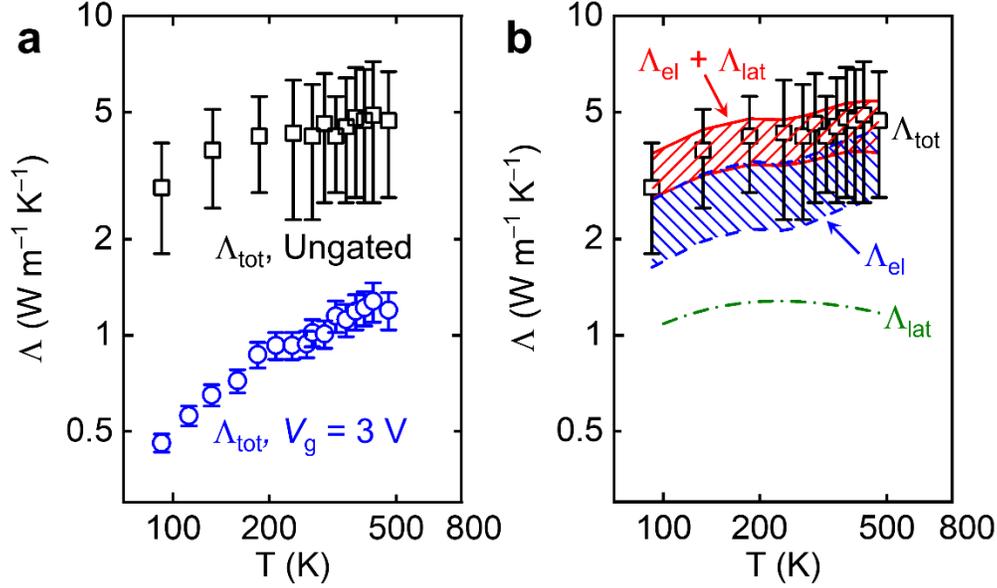

**Fig. 4 | *T*-dependent thermal conductivity. a** *T*-dependent thermal conductivities of P-LSCO (ungated, black squares) and BM-LSCO ($V_g = 3$ V, blue circles). **b** *T*-dependent $\Lambda_{el}$ and $\Lambda_{lat}$ of P-LSCO. BTE calculations were performed to predict $\Lambda_{lat}$ for a 45-nm $La_{0.5}Sr_{0.5}CoO_{3-\delta}$ ($\delta = 0.1$) film (green dashed-dotted line), while $\Lambda_{el}$ (blue dashed line) was calculated from the temperature dependence of the resistivity (Fig. 1d) using the WFL with the common lower and upper limits of *L*. Red solid lines represent the sum of $\Lambda_{el}$ and $\Lambda_{lat}$.

In contrast to P-LSCO, BM-LSCO is strongly insulating, with lattice vibrations therefore being the dominant heat carriers, and thus $\Lambda_{tot} \approx \Lambda_{lat}$. However, $\Lambda_{lat}$ of BM-LSCO still does not show a peak in the probed temperature range, in contrast to typical behavior of pristine insulating crystals. The reason for this is that BM-LSCO contains an even larger density of $V_{OS}$, a lattice structure with a lower symmetry due to those $V_{OS}$, and additional disorder associated with the A-site cations (La/Sr), resulting in strong extrinsic phonon scattering, even at low *T*. Therefore, the thermal transport in BM-LSCO is more glass-like, rather than crystal-like. Supplementary Fig. 8 also shows the $T$-dependent thermal conductivity of one BM-LSCO sample gated to 4 V, which exhibits a similar trend to that of the BM-LSCO sample at 3 V shown in Fig. 4a, confirming these results and conclusions.

## Conclusion

In summary, we have demonstrated that by applying single-step solid-state electrolyte gating at RT we can achieve a record-high (> 5-fold) tuning of the thermal conductivity of ion-gel-gated LSCO films in a non-volatile, continuous manner. This large tunability in the thermal conductivity of epitaxial LSCO films results from the structural modifications and metal-insulator transition induced by the P → BM phase transformation. As $V_g$ increases, more oxygen vacancies are created and the lattice symmetry is lowered, leading to a sizable reduction in the LSCO lattice thermal conductivity. Accompanying these structural modifications, the metal-insulator transition across the P → BM transformation further extends the extent of the tuning of thermal conductivity by controlling the electronic contributions to the thermal conductivity. In addition, $T$-dependent thermal conductivity measurements provided further physical insights into the scattering mechanisms of the relevant heat carriers. For metallic P-LSCO, the observed increasing trend of $\Lambda_{tot}$ *vs.* $T$ is partially attributed to the substantial electronic contributions to thermal transport, which are absent for BM-LSCO. For insulating BM-LSCO, the increase in thermal conductivity with $T$ instead represents more of the trend typical of amorphous oxides due to the extensive structural defects (*e.g.*, higher $V_O$ concentrations and A-site disorder). This success in actively tailoring the thermal conductivity of thin-film LSCO by over a factor of five, *via* electrolyte gating, opens the possibility of voltage-driven tunable thermal materials for applications in thermal management and energy conversion, which require dynamic control of heat propagation.

## Methods

**LSCO film growth, device fabrication, and electrolyte gating.** Thin films of LSCO were deposited on commercial (MTI corp.) $5 \times 5$ mm$^2$ and $10 \times 10$ mm$^2$ LaAlO$_3$(001) substrates using high-pressure-oxygen sputtering, from 2" polycrystalline sputtering targets of the same nominal stoichiometry. Substrates were first annealed at 900°C in flowing O$_2$ (99.998%, ~1 Torr) for 15 min, before being cooled to 600°C for deposition. LSCO was then deposited *via* DC sputtering at 600°C in flowing O$_2$ (1.5 Torr) at a power of $60 - 70$ W. This produced epitaxial films at a deposition rate of ~20 Å/min. After deposition, films were cooled to RT in 600 Torr of O$_2$ at ~15 °C/min. These procedures essentially employ optimized parameters reported in prior studies[6,14-16,30,37]. The thicknesses of films in this work varied from 45 to 58 nm, estimated from established growth rates and cross-checked from wide-angle XRD Laue fringe spacings.

Sputtered LSCO films were then used to fabricate electrolyte-gate transistor devices, as also reported previously[6,14,15]. As-deposited LSCO films were first Ar-ion milled selectively using steel masks to form LSCO channels of $1 \times 1$ mm$^2$. Subsequent Mg(5 nm)/Pt(50 nm) gate and contact electrodes were then sputter-deposited through a separate mask and annealed in O$_2$ (450°C, 5 min). To complete side-gated transistors, a single piece of ion gel was then cut and laminated atop the LSCO channel, contact electrodes (partially), and gate electrodes (substantially). A schematic of the final device is shown in Fig. 1a. The ion gels were fabricated by spin coating (to ~50 μm thickness) a solution of ionic liquid and polymer onto ~1 inch$^2$ glass wafers. The ionic liquid solution consisted of: 1-ethyl-3-methylimidazolium bis(trifluoro-methylsulfonyl)imide (EMI-TFSI) ionic liquid, poly(vinylidene fluoride-cohexafluoropropylene) polymer, and acetone, in a weight ratio of 1:4:8, respectively. Gating was performed at RT, in vacuum ($<1 \times 10^{-5}$ Torr), sweeping $V_g$ at 0.5 mV sec$^{-1}$ to the specified target values; $V_g$ was applied between the side gate pads and two diagonal electrodes shorted to the film channel (see Fig. 1a). During $V_g$ sweeps, two-wire resistance measurements were made *in situ* between two diagonal electrodes (the two contacts that were not shorted to the gate counter-electrode), representing a source and drain. Such measurements were made by applying a voltage ($V_{SD}$) of ±0.2 V and measuring a current ($I_{SD}$) (see Supplementary Fig. 2). $V_g$ and $V_{SD}$ were applied with separate Keithley 2400 (K2400) source-measure units, while the gate current ($I_g$) and $I_{SD}$ were measured with the corresponding K2400 units. After reaching the desired $V_g$, gating was terminated by disconnecting the voltage supply to the gates, removing the ion gel, and cleaning the film surface with acetone to remove any residual ion gel.

**X-ray diffraction.** LSCO film devices of two dimensions ($1 \times 1$ mm$^2$ films on $5 \times 5$ mm$^2$ substrates, and $4 \times 3.5$ mm$^2$ films on $10 \times 10$ mm$^2$ substrates) were characterized both before and after gating *via* high-resolution specular wide-angle XRD (using a Rigaku Smartlab XE, with Cu Kα radiation). Reciprocal space mapping was performed with the same instrument and settings. Synchrotron XRD measurements (shown only in Supplementary Fig. 1a) were performed on a representative $4 \times 3.5$ mm$^2$ LSCO film deposited on a $10 \times 10$ mm$^2$ LAO(001) substrate at the 12-ID-D beamline of the Advanced Photon Source at Argonne National Lab. The synchrotron XRD set-up was equipped with a six-circle Huber goniometer and a Pilatus II 100 K area detector. The spot size and X-ray beam energy were ~500 $\mu$m and 22 keV ($\lambda \sim 0.56$ Å), respectively. The measurements were carried out at a temperature of 150 K with a liquid-N$_2$ gas flow cryocooler.

**Electronic transport.** Four-wire electronic transport measurements were taken in the van der Pauw geometry before and after electrolyte gating of LSCO films (*ex situ*), to determine the LSCO channel resistivity. Temperature-dependent data were acquired in a Quantum Design Physical Property Measurement System (PPMS). Electronic transport measurements were taken using either quasi-AC with the PPMS internal bridge (for metallic samples), or DC with a Keithley 2400 current source and a Keithley 2002 multimeter (for insulating samples).

**Determination of V$_O$ concentrations, $\delta$.** The oxygen deficiency, $\delta$, of P-phase LSCO films was estimated using a previously established and validated method[14], in which the measured film resistivity is compared to La$_{1-x}$Sr$_x$CoO$_3$ ($x \leq 0.30$) single crystal resistivity data (where $\delta$ is close enough to zero to be considered negligible) to estimate the film $\delta$. First, the known low-temperature single crystal resistivity is plotted *vs. x*, then the measured film resistivity is used to interpolate an effective $x$ value, $x_{eff}$, from this master curve. Using $x_{eff} = x - 2\delta$ (*i.e.*, assuming each V$_O$ dopes 2 electrons), with $x = 0.5$ in our case, the film $\delta$ can then be estimated. This estimation can be made only in the P phase, and only up to $\delta = 0.25$. We do apply this method in the mixed phase region, where the current is expected to be shunted by P regions, but it cannot, therefore, be applied in the BM phase. As noted in connection with Fig. 3c, at $V_g = 3$, 3.5 and 4 V we simply assume $\delta \approx 0.5$ based on the BM structure evident from XRD.

**TDTR measurements.** The thermal conductivities of LSCO films were measured with TDTR[49]. Prior to TDTR thermal measurements, a ~70 nm Pt layer was sputter-deposited onto the LSCO films (after gating to the desired $V_g$), to act as a transducer layer for TDTR measurements. At the laser wavelength of 783 nm, the extinction coefficient is ~7.4 for Pt[69]; therefore, the thickness of 70 nm is sufficient for the Pt transducer to be considered optically opaque. A Si/SiO$_2$ substrate was also coated with Pt in the same deposition, providing a reference to cross-check the electrical and thermal properties of the Pt. In TDTR, a mode-locked Ti:sapphire laser produces a train of optical pulses (~100 fs in duration) at a repetition rate of 80 MHz and a central wavelength of ~780 nm. The laser is divided into a pump beam and a probe beam with two orthogonal polarizations by a polarizing beam splitter. The pump beam is modulated by an electro-optical modulator, which heats the sample. The probe beam is modulated by a mechanical chopper and detects the temperature response of the sample upon pump heating. An objective lens is used to focus the pump and probe beams onto the sample surface. A mechanical delay stage then varies the optical path of the pump beam, which produces a time delay of up to 4 ns between pump heating and probe sensing. The probe beam reflected from the sample is collected with a fast-response Si detector and then amplified by an RF lock-in amplifier for the data reduction.

We used a 5× objective lens with a $1/e^2$ beam spot size of ≈12 μm for all samples. For temperature-dependent TDTR measurements over the range of ~90 to ~500 K, the sample was mounted on a heating/cooling stage. Specifically, measurements at low temperatures (from ~90 K to RT) were done under vacuum, while measurements at elevated temperatures (from RT to ~500 K) were carried out in air. A MK2000 series temperature controller was used to provide temperature control. The set temperature of the controller was defined as the setting temperature. The laser power was optimized as a compromise between the signal-to-noise ratio and steady-state heating ($\Delta T_s$) for each set temperature. For measurements at low temperatures, due to the significant decrease in heat capacity, the temperature rise induced by the pump per-pulse can also be large ($\Delta T_p$). Therefore, we performed a temperature correction for low-temperature measurements, taking into account both $\Delta T_s$ and $\Delta T_p$. The detailed procedures for this temperature correction are provided in Supplementary Section 6.

## Data availability

All data needed to evaluate the conclusions in this paper are present in the paper and/or the Supplementary Information. Additional data related to this work may be requested from the corresponding authors.

## Code availability

All codes related to this work are available from the corresponding authors upon reasonable request.

## Acknowledgements

This work was primarily supported by the National Science Foundation (NSF) through the UMN MRSEC under DMR-2011401. Parts of this work were carried out at the Characterization Facility, UMN, which receives partial support from the NSF through the MRSEC program. Portions of this work were also conducted in the Minnesota Nano Center, which is supported by the NSF through the National Nanotechnology Coordinated Infrastructure under ECCS-2025124. This research used resources of the Advanced Photon Source, a U.S. Department of Energy (DOE) Office of Science user facility operated for the DOE Office of Science by Argonne National Laboratory under Contract No. DE-AC02-06CH11357. R.X., H.Z., and T.F. acknowledge the support from NSF (under CBET-2212830). The computation used resources of the National Energy Research Scientific Computing Center, supported by the Office of Science of the DOE (under Contract DE-AC02-05CH11231), the Center for High Performance Computing at the University of Utah, and the Extreme Science and Engineering Discovery Environment.

## Author contributions

C.L. and X.W. originated and supervised the research. Y.Z. and C.Z. carried out the TDTR measurements and analyses under the guidance of X.W.. W.M.P., V.C., and K.H. prepared the LSCO samples under the guidance of C.L.. W.M.P., V.C., and H.Z. performed the XRD characterization. W.M.P. fabricated the devices and performed the electrolyte gating and electrical transport measurements and analyses, under the guidance of C.L.. R.X., H.Z., and T.F. did the MD simulations and BTE calculations. Y.Z., W.M.P., R.X., T.F., C.L., and X.W. contributed to the writing of the manuscript. All authors discussed the data and provided input on the paper.

## Competing interests

The authors declare no competing interests.

## Additional information

Supplementary information. The online version contains supplementary material available at [to be inserted by the editor].

# Wide-range continuous tuning of the thermal conductivity of La$_{0.5}$Sr$_{0.5}$CoO$_{3\text{-}\delta}$ films *via* room-temperature ion-gel gating

Yingying Zhang[1], William M. Postiglione[2], Rui Xie[3], Chi Zhang[1], Hao Zhou[3], Vipul Chaturvedi[2], Kei Heltemes[2], Hua Zhou[4], Tianli Feng[3], Chris Leighton[2*] and Xiaojia Wang[1*]

[1]*Department of Mechanical Engineering, University of Minnesota, Minneapolis, MN55455, USA*

[2]*Department of Chemical Engineering and Materials Science, University of Minnesota, Minneapolis, MN55455, USA*

[3]*Department of Mechanical Engineering, University of Utah, Salt Lake City, Utah 84112, USA*

[4]*Advanced Photon Source, Argonne National Laboratory, Lemont, Illinois 60439, USA*

## 1. Structural characteristics of LSCO films

Details for samples measured in this study are summarized in Table S1, including the channel area, film thickness, gate voltage, crystalline phase, and out-of-plane lattice parameter.

**Table S1.** Summary of film information for XRD and TDTR measurements corresponding to Fig. 1c and Fig. 3a. *Asterisks indicate films for which XRD data were not taken, and therefore phase composition is inferred based on electronic transport, TDTR, and XRD from other samples.

| Sample # | Channel area (mm$^2$) | Film thickness (nm) | Gate voltage (V) | Phase | $c$-axis lattice parameter (Å) |
|---|---|---|---|---|---|
| 1 | $1 \times 1$ | 45 | 0.0 | P | 3.88 |
| 2 | $1 \times 1$ | 45 | 1.0 | P | 3.89 |
| 3 | $1 \times 1$ | 45 | 1.4 | P + BM* | --- |
| 4 | $1 \times 1$ | 45 | 1.8 | P + BM* | --- |
| 5 | $1 \times 1$ | 45 | 2.0 | P + BM | 3.94 |
| 6 | $1 \times 1$ | 45 | 2.5 | P + BM* | --- |
| 7 | $1 \times 1$ | 45 | 2.6 | P + BM* | --- |
| 8 | $1 \times 1$ | 45 | 3.0 | BM | 3.99 |
| 9 | $1 \times 1$ | 58 | 4.0 | BM | 4.00 |
| 10 | $4 \times 3.5$ | 45 | 3.5 | BM | 3.99 |

*Authors to whom correspondence should be addressed. Electronic mail: leighton@umn.edu; wang4940@umn.edu

In addition to the structural characterization shown in Fig. 1c of the main text, further characterization of as-grown perovskite-phase La$_{0.5}$Sr$_{0.5}$CoO$_{3-\delta}$ (P-LSCO) films were performed *via* wide-angle specular XRD using synchrotron radiation ($\lambda = 0.5636$ Å) at the Advanced Photon Source, Argonne National Lab, and reciprocal space mapping (RSM) using in-house Cu K$\alpha$ radiation (see Methods). Figure S1a shows the synchrotron XRD pattern collected around the substrate LAO(002) reflection for an as-grown P-LSCO film ($4 \times 3.5$ mm$^2$), this film being representative of other LSCO films investigated in this work. The visible Laue oscillations in Figure S1a indicate low surface and substrate/interface roughnesses, the oscillation periodic spacing indicating a thickness of ~45 nm. The P-LSCO (002) peak was fit to a Gaussian, and the extracted full-width-at-half-maximum produced a coherence length comparable to the film thickness (~35 nm), further indicative of high crystalline quality. Figure S1b additionally shows the reciprocal space map (RSM) of a similar film about the LAO(103) asymmetric reflection. The P-LSCO film intensity is centered along the vertical line $H = 1$, indicating that the film has the same in-plane lattice parameter as the substrate, *i.e.*, it is fully strained, with some minor strain relaxation evidenced by the broad, low-intensity region skewed toward the relaxed position. Note that the (103) Bragg reflection peak for the LAO substrate is relatively broad due to the twin domains prevalent in commercial LAO substrates.

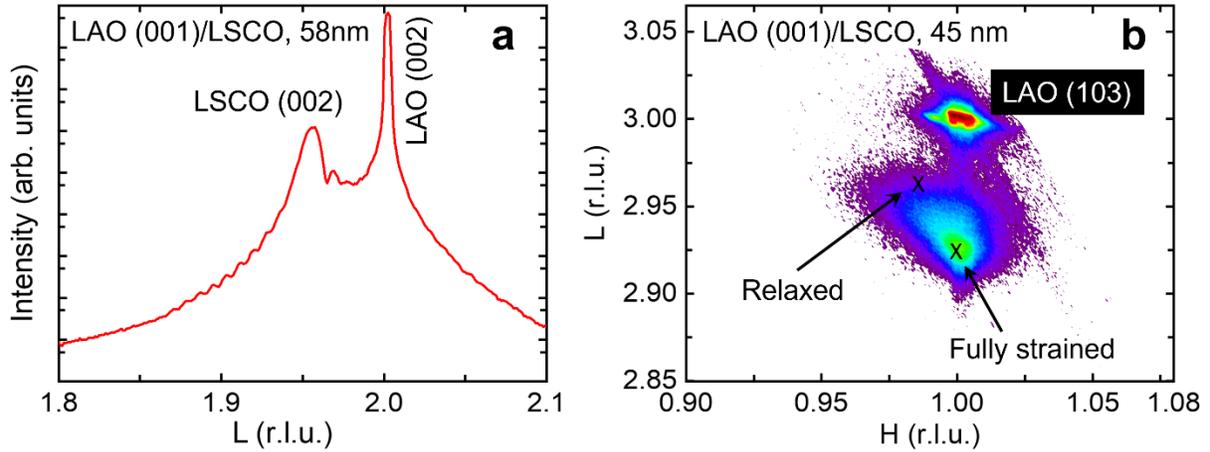

**Fig. S1 | Wide-angle specular X-ray diffraction characterization of as-grown P-LSCO films sputter-deposited on LAO(001) substrates**. **a** Specular (002) synchrotron X-ray diffraction ($\lambda = 0.5636$ Å) peaks of the 45-nm P-LSCO film and substrate. **b** Reciprocal space map taken around the LSCO film and LAO substrate (103) peaks. The ×'s mark the positions of the film peak expected for the fully strained (pseudomorphic) and relaxed positions, respectively. "r.l.u." (reciprocal lattice units) are with respect to the LAO substrate ($a = 3.79$ Å).

XRD characterization was also performed on gated films, as shown in Fig. 1c in the main article. Scans on the samples gated to 1, 2, and 3 V have relatively low signal-to-noise ratio, especially around the emerging BM peaks (006) and (0010), due to the small film area ($1 \times 1$ mm$^2$) in this case. To more carefully probe the BM structure, a larger-channel-area device ($4 \times 3.5$ mm$^2$) was also gated to 3.5 V, and the XRD results are shown in Fig. 1c (ungated (black) and 3.5 V (magenta)). The more optimized geometry in this sample reduces the intensity of the 'ungated' perovskite (*i.e.*, the film buried beneath the Pt electrode), reduces the intensity of the Pt(111) peak (from the Pt electrodes), and significantly increases the signal-to-noise ratio for the LSCO film peaks, which is especially important for the lower-intensity BM (006) and (0010) reflections. This improved signal-to-noise ratio allows more quantitative analysis, where the full-width-at-half-maximum (FWHM) was estimated by using Gaussian peak fitting, for example. While the positions of the BM (006) and (0010) peaks gave the same out-of-plane lattice parameter as the (008) (~4.0 Å), suggesting all peaks indeed derive from the same phase, the (006) and (0010) are significantly broader (larger FWHM), resulting in lower out-of-plane coherence lengths (as calculated from the Scherrer formula). The extracted coherence length was found to be ~$10 - 15$ nm for the (006) and (0010) *versus* $25 - 30$ nm for the (008), indicating substantial disorder in the BM phase. Additionally, the observed intensity ratio between BM (008) and (006) (or (0010)) was ~70, while the ratio expected for fully ordered BM is ~$10 - 40$, as found in various reports of high-quality *as-grown* BM SrCoO$_{2.5}$ films[1-6]. This larger intensity ratio is consistent with prior work for ion-gel-gate-induced BM-LSCO[7], and is likely due to the presence of additional defects which disrupt the long range V$_O$ ordering. Based on the above observations, we thus conclude that the BM-LSCO formed here *via* ion-gel-gating, particularly the anion (oxygen) sublattice, features significant additional disorder compared to as-grown BM SCO films.

## 2. Electrolyte gating: *in-situ* two-terminal resistance *vs.* gate voltage

Ion-gel gating of 45-58 nm LSCO films was carried out at 300 K under vacuum ($< 1 \times 10^{-5}$ Torr) by sweeping the gate voltage ($V_g$) at a rate of 0.5 mV sec$^{-1}$ to the target voltage (see Methods). *In-situ* electronic transport measurements were taken by applying a source-drain voltage ($V_{SD} = \pm$ 0.2 V) throughout the experiment, measuring the source-drain current, and determining the source-drain resistance, $R_{SD}$, as a function of $V_g$. Figure S2 shows the $R_{SD}$ *vs.* $V_g$ measured for the 11 $1 \times$ 1 mm channel films in Table S1. Gating starts at $V_g = 0$, sweeping to larger positive $V_g$ at 0.5 mV sec$^{-1}$ and stopping at specific $V_g$ values as needed for *ex situ* characterization (XRD, resistivity, and TDTR). The films start as perovskite (P) ($R_{SD}$ ~ 200 $\Omega$) and gradually increase in resistance with increasing $V_g$ until the transformation to brownmillerite begins. Then a rapid increase in $R_{SD}$ occurs on entry into the phase coexistence region (P + BM) around $2 - 3$ V. This is followed by another relatively flat region ($R_{SD}$ ~ $10^5$ $\Omega$) indicating fully formed BM. Overall, the ion-gel-gated samples show good repeatability, with only minor sample-to-sample variations, around the transition region (2 − 3 V) and at the highest $V_g$ (4 V). Note that the 4 × 3.5 mm channel sample is not shown here due to the larger resistance owing to its larger channel length. Finally, after the conclusion of each gating experiment, the ion gels were removed, the films cleaned with acetone, and *ex-situ* four-wire van der Pauw resistance measurements were performed (see Fig. 1d & Fig. 3a in the main article).

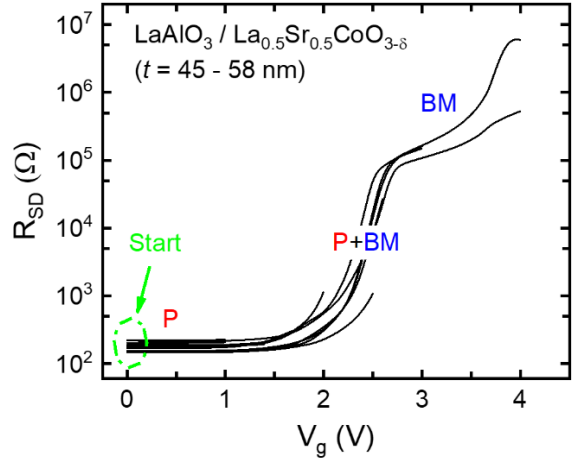

**Fig. S2 | Two-terminal resistance.** Source-drain two-terminal resistance ($R_{SD}$) *vs.* gate voltage ($V_g$) for 11 1 × 1 mm channel area ion-gel-gated LSCO films of 45-58 nm thickness on LAO(001) substrates. Gating was performed at 300 K under vacuum.

## 3. Interfacial thermal conductance of Pt/LSCO and LSCO/LAO interfaces

In TDTR measurements, only parameters with sufficient sensitivities can be uniquely determined. The measurement sensitivity to a certain parameter is calculated as[8]:

$$S_\alpha = \frac{\partial \ln R}{\partial \ln \alpha},$$ (1)

where $\alpha$ is the parameter of interest, $S_\alpha$ is the measurement sensitivity to the parameter $\alpha$, and $R$ is the ratio signal from TDTR. In TDTR measurements, heat propagation into the sample is treated as thermal waves at the modulation frequency $f$. Subsequently, the thermal penetration depth of the thermal wave ($h$) can be calculated based on $h = \sqrt{\frac{\Lambda}{\pi C f}}$, where $C$ is the heat capacity and $\Lambda$ is the thermal conductivity of the sample material. For thermal measurements of thin films, when $h$ is smaller than the film thickness $d$, the film is thermally opaque. The TDTR signal is then not sensitive to the substrate and the interfacial thermal conductance of the interface between the film and the substrate ($G_2$). On the contrary, when $h$ is larger than $d$, the film becomes thermally thin. In this case, the TDTR signal becomes sensitive to the properties of the LAO substrate ($\Lambda_{LAO}$ and $C_{LAO}$) and $G_2$. Due to the ultra-thin film nature of our LSCO samples, $h$ is larger than the film thickness for all three modulation frequencies used in TDTR measurements (1.5, 9.0, and 18.8 MHz). Therefore, measurement conditions must be carefully designed to distinguish the sensitivities to $\Lambda_{LSCO}$, $G_1$, and $G_2$ for reliable determination of these parameters.

Here, we take the as-grown P-LSCO sample (ungated) as an example to show how we design the experiments and data reductions to extract $\Lambda_{LSCO}$, $G_1$, and $G_2$ with reasonable accuracy. Figure S3a-c shows the routine TDTR measurement sensitivity to multiple parameters under different modulation frequencies for the ungated sample. It is found that routine TDTR measurements with three modulation frequencies have the highest sensitivities to the properties of the Pt transducer and LAO substrate. Therefore, prior to measuring the LSCO films, we obtained the properties of the Pt transducer and LAO substrate by measuring reference samples of Pt on Si/SiO$_2$(300 nm) and Pt on LAO prepared from the same sputtering batch. These properties of the transducer and substrate were used as input parameters for data reduction of LSCO thin-film measurements to improve the reliability. For routine TDTR, the measurement sensitivity is more reasonable for $\Lambda_{LSCO}$ than $G_1$ or $G_2$. In this case, we extracted $\Lambda_{LSCO}$ from routine TDTR measurements by simultaneously fitting all three modulation frequencies. For determination of $G_2$, since routine TDTR measurements have low sensitivity to $G_2$, the dual-frequency TDTR approach was adopted to increase the sensitivity to $G_2$[9]. As illustrated in Fig. S3d, the sensitivities of dual-frequency TDTR measurements (18.8 MHz/1.5 MHz) to the Pt transducer and LAO substrate are largely suppressed, while the sensitivity to $G_2$ is improved. Using the dual-frequency approach, $G_2$ was determined to be $800 \pm 500$ MW m$^{-2}$ K$^{-1}$, which agrees with the literature values for strongly bonded interfaces[10]. The ~60% uncertainty on $G_2$ does not propagate significant errors into the overall uncertainty of $\Lambda_{LSCO}$. This is because the routine TDTR measurement sensitivity to $G_2$ is much lower than the measurement sensitivity to $\Lambda_{LSCO}$.

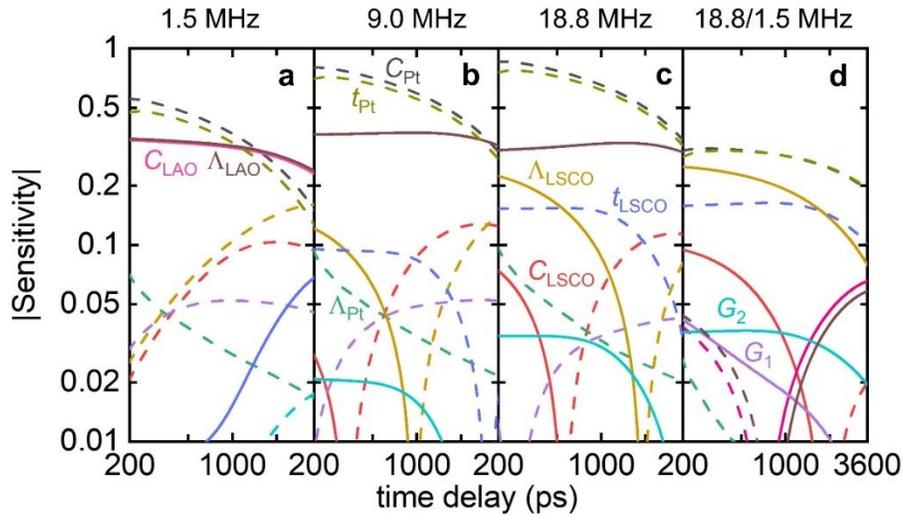

**Fig. S3 | Representative sensitivity analyses.** Sensitivity plots of TDTR measurements for the ungated P-LSCO sample calculated with the following parameters: heat capacities ($C_{Pt}$, $C_{LSCO}$, $C_{LAO}$), film thicknesses ($t_{Pt}$, $t_{LSCO}$), thermal conductivities ($\Lambda_{Pt}$, $\Lambda_{LSCO}$, $\Lambda_{LAO}$), interfacial thermal conductances ($G_1$ and $G_2$), and the beam spot size of 12 μm. Panels **a-c** are routine TDTR ratio sensitivity to multiple parameters with different modulation frequencies. Panel **d** is the dual-frequency measurement sensitivity. Solid lines mean the sensitivity value is positive, while dashed lines represent negative sensitivity values.

Also from Fig. S3, the measurement sensitivity to $G_1$ is low in all cases due to the large value of $G_1$ of the ungated sample, leading to large measurement uncertainties. This is expected considering the extra thermal transport channel at the Pt/P-LSCO interface through electron-electron coupling. When $G_1$ decreases as P-LSCO (metallic) transforms to BM-LSCO (insulating) with electrolyte gating, the measurement sensitivity to $G_1$ is improved, which allows for the extraction of $G_1$ for samples gated at higher voltages ($V_g > 2$ V, see Fig. 3b). To validate our measured values of $G_1$ for BM-LSCO films exhibiting insulting behavior, we list $G$ values for interfaces between Pt and different insulating oxides[11-14] in Fig. S4, which vary from 180 to 330 MW m$^{-2}$ K$^{-1}$. Our measurement results for $G_1$ between Pt and BM-LSCO samples (gated at 3, 3.5 and 4 V with negligible electronic contribution) are determined to be 160, 150 and 200 MW m$^{-2}$ K$^{-1}$, respectively, in good agreement with literature data.

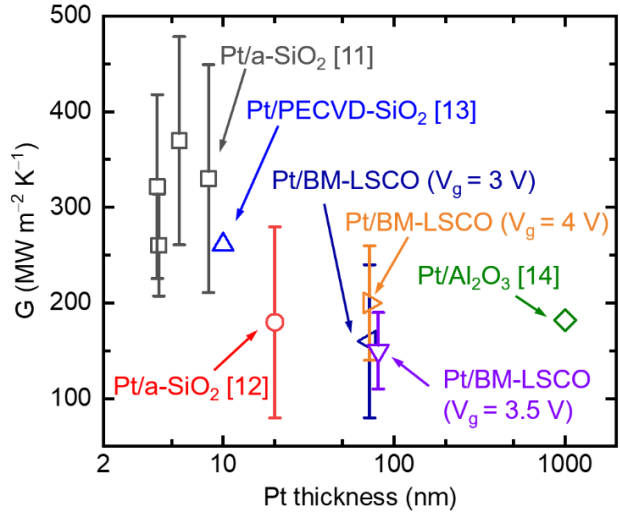

**Fig. S4 | Interfacial thermal conductance.** Summary of the interfacial thermal conductance between Pt and oxides plotted against the Pt thickness. The $G$ values of Pt/SiO$_2$ and Pt/Al$_2$O$_3$ are taken from the literature[11-14].

For samples gated at $V_g \leq 2$ V, with larger $G_1$ values, a nominal value of 400 MW m$^{-2}$ K$^{-1}$ was used for the data reduction. Due to the low sensitivity to $G_1$, the determination of $\Lambda_{LSCO}$ is not significantly impacted by $G_1$. Table S2 summaries the values of $G_1$ and $\Lambda_{LSCO}$ from measurement fitting. When both parameters are fitted simultaneously, $G_1$ varies from 250 to 920 MW m$^{-2}$ K$^{-1}$, while $\Lambda_{\text{LSCO}}$ only varies from 4.1 to 5.5 W m$^{-1}$ K$^{-1}$. When $G_1$ is fixed at 400 MW m$^{-2}$ K$^{-1}$, the fitted $\Lambda_{\text{LSCO}}$ (listed in the last column of Table S2) varies from 4.1 to 6.3 W m$^{-2}$ K$^{-1}$. The variations of $\Lambda_{\text{LSCO}}$ with or without $G_1$ being fixed in the data analysis are all within the measurement uncertainty.

**Table S2.** Results of $G_1$ and $\Lambda_{\text{LSCO}}$ for the ungated samples measured at different locations and modulation frequencies, obtained from simultaneous fitting for both $G_1$ and $\Lambda_{\text{LSCO}}$ or fitting for $\Lambda_{\text{LSCO}}$ only ($G_1 = 400$ MW m$^{-2}$ K$^{-1}$).

| Sample # | Location # | Modulation frequency (MHz) | $G_1$ (MW m$^{-2}$ K$^{-1}$) | $\Lambda_{\text{LSCO}}$ (W m$^{-1}$ K$^{-1}$) | $G_1$ (MW m$^{-2}$ K$^{-1}$) | $\Lambda_{\text{LSCO}}$ (W m$^{-1}$ K$^{-1}$) |
|---|---|---|---|---|---|---|
| **1** | L1 | 1.5 | 920 | 4.7 | | 6.3 |
| | | 9 | 490 | 4.3 | | 4.3 |
| | | 18.8 | 240 | 5.0 | | 5.1 |
| | L2 | 1.5 | 280 | 5.5 | | 4.5 |
| | | 9 | 530 | 4.2 | | 4.1 |
| | | 18.8 | 320 | 4.6 | | 4.7 |
| | L3 | 1.5 | 390 | 4.6 | | 4.5 |
| | | 9 | 600 | 4.5 | 400 | 4.3 |
| | | 18.8 | 410 | 4.1 | | 4.1 |
| **2** | L1 | 1.5 | 520 | 4.3 | | 4.9 |
| | | 9 | 660 | 4.1 | | 4.3 |
| | | 18 | 330 | 4.9 | | 5.0 |
| | L2 | 1.5 | 440 | 4.7 | | 4.9 |
| | | 9 | 250 | 4.9 | | 4.4 |
| | | 18 | 440 | 4.5 | | 4.5 |
| **Average (range)** | | | 455 (240 - 920) | 4.6 (4.1 − 5.5) | 400 (fixed) | 4.7 (4.1 − 6.3) |

In addition, we followed the thickness-extrapolation method used by Wu *et al.* for quantifying the lumped thermal resistances of two interfaces ($R_{\text{eff}} = 1/\Lambda_{\text{LSCO}} + 1/G_1 + 1/G_2$)[15], as a criterion to cross check the reasonableness of our measured $G_1$ and $G_2$. We prepared ungated LSCO films with varying thicknesses and coated them with Pt. The lumped thermal resistances $R_{\text{eff}}$ of Pt/P-LSCO measured with TDTR are represented in Table S3 as a function of the LSCO film thickness. The linear extrapolation of $R_{\text{eff}}$ to zero film thickness (intercept on the $y$-axis) gives $1/G_1$ $+ 1/G_2 = 6.15$ nm m K W$^{-1}$ with ~50% uncertainty. With $G_1 = 400$ MW m$^{-2}$ K$^{-1}$ and $G_2 = 800$ MW m$^{-2}$ K$^{-1}$, the calculated $R_{\text{eff}}$ value (3.75 nm m K W$^{-1}$) is within the uncertainty.

**Table S3.** $R_{\text{eff}}$ of ungated LSCO films with varying thicknesses on LAO substrates.

| Film thickness (nm) | Lumped thermal resistance $R_{\text{eff}}$ (nm m K W$^{-1}$) |
|---|---|
| 5.2 | 6.42 ± 2.24 |
| 10.7 | 9.17 ± 2.75 |
| 15.6 | 9.36 ± 2.81 |
| 19.8 | 8.59 ± 2.83 |

## 4. Uncertainty analysis for TDTR measurements

For TDTR measurements, the uncertainty on LSCO thermal conductivity is calculated based on the sensitivity analysis using the following expression:

$$\left(\frac{\Delta\Lambda}{\Lambda}\right)^2 = \sum \left(\frac{S_\alpha}{S_\Lambda}\frac{\Delta\alpha}{\alpha}\right)^2 + \left(\frac{S_\phi}{S_\Lambda}\delta\phi\right)^2, \qquad (2)$$

where $\Delta\Lambda/\Lambda$ is the relative uncertainty of the sample thermal conductivity $\Lambda$, $\Delta\alpha/\alpha$ is the relative uncertainty of a certain parameter $\alpha$, $\delta\phi$ is the uncertainty of the phase, and $S_\alpha$ is the measurement sensitivity to a parameter $\alpha$. With a combination of a 3% uncertainty in the beam spot size, a 4% uncertainty in the transducer thickness, a 3% uncertainty in the heat capacities of the transducer, sample, and substrate, and an 11% uncertainty in the substrate thermal conductivity, the resulting uncertainty on the ungated P-LSCO thermal conductivity is ~43%. The main contributions to the total uncertainty on the LSCO thermal conductivity are from $\Lambda_{\text{LAO}}$, $C_{\text{LAO}}$, $d_{\text{Pt}}$ and $C_{\text{Pt}}$. As $\Lambda_{\text{LSCO}}$ decreases when $V_g$ increases, the measurement sensitivity to $\Lambda_{\text{LSCO}}$ is also improved. As shown in Fig. S5, the measurement sensitivity to $\Lambda_{\text{LSCO}}$ for the BM-LSCO sample gated at 3 V significantly increases compared to the ungated P-LSCO sample. Therefore, the uncertainty of the 3-V-gated LSCO is reduced to ~10%.

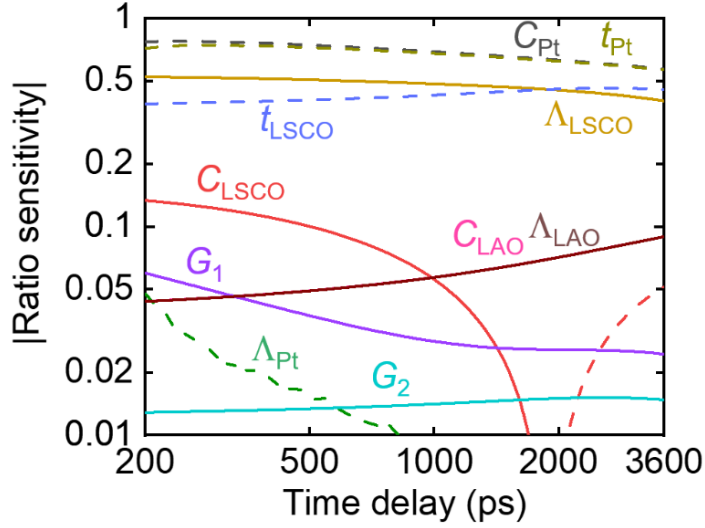

**Fig. S5 | Sensitivity plot for uncertainty analyses**. Sensitivity plot for routine TDTR measurements at 18.8 MHz for the LSCO sample gated at 3 V for uncertainty calculations.

## 5. MD simulation of LSCO thermal conductivity

To study the effect of oxygen defects on thermal transport in LSCO, we theoretically predicted the lattice thermal conductivity of LSCO with different oxygen deficiencies ($\delta = 0.1$, 0.25, 0.375, and 0.5) using molecular dynamics (MD) simulations (Fig. 3d) with the LAMMPS package. The potentials used in LAMMPS in this work are the short-range Buckingham potential + long-range Coulombic potential, which are two-body potentials. Therefore, we expect that the heat flux formalism in LAMMPS should be reliable. Classical potentials are not as accurate as first-principles calculations, nor can they correctly predict optical branches. However, our goal here is to provide qualitative insights to better interpret the experimental results, rather than pursuing very high accuracy in BTE calculations. The equilibrium Green-Kubo method was used to calculate the thermal conductivity of LSCO:

$$k_i = \frac{1}{k_{\mathrm{B}}T^2 V} \int_0^\infty < S_i(t) \cdot S_i(0) > dt, i = x, y, z, \tag{3}$$

where $S$ is the transient heat flux, and $<S_i(t) \cdot S_i(0)>$ is the heat flux autocorrelation function, which measures the correlation between the heat flux at any time with itself after time $t$. $k_B$ is the Boltzmann constant, $T$ is temperature, and $V$ is the volume of the simulation domain. The domain was constructed by using $12 \times 12 \times 12$ unit cells (8640 atoms) of $LaCoO_3$ and replacing half the La atoms with Sr. Oxygen atoms were removed to create point vacancy defects either randomly or periodically depending on the needs. Our algorithm ensures the vacancies are point defects instead of cluster defects. The Buckingham and Coulomb interatomic potential was used to describe the interactions[16]. The timestep was set as 0.5 fs. We first relaxed the geometries at constant pressure and temperature (NPT ensemble) at 2 K for 0.2 ns (400000 steps) and then increased the temperature to 300 K and ran for another 0.2 ns. Then, the ensemble was changed to constant volume and energy (NVE ensemble) and run for 0.2 ns to stabilize the ensemble. Finally, the NVE ensemble was run for another 2 ns to compute the heat flux and auto-correlation function. For each individual structure, the thermal conductivity was obtained by averaging the thermal conductivities of three independently generated structures with randomly distributed $V_O$ along the $x$, $y$, and $z$ directions. In addition, for the case of $\delta = 0.5$, we generated two sets of structures: disordered $V_O$ and ordered $V_O$, with the latter case being the orthorhombic BM phase.

To validate the equilibrium GKMD results, we further performed non-equilibrium molecular dynamics (NEMD) for four cases with the non-stoichiometry $\delta$ values that have been studied with EMD (*i.e.*, $\delta = 0.1$, 0.25, 0.375, and 0.5). The NEMD simulation domain is shown in Fig. S6a with a size of $2.2938 \times 3.0584 \times 3.0584$ nm$^3$. NEMD simulations were performed using the LAMMPS package with the same classical potential settings as implemented in the GKMD. The hot reservoir was set in the middle of the simulation domain with a length of 2 nm, and the cold reservoir of the same size was split into two pieces equally and placed at the two edges of the box along the *x* direction. In all the three dimensions, periodic boundary conditions were applied. A Langevin thermostat was adopted to maintain the temperatures of the hot reservoir at 598.15 K and the cold reservoir at 548.15 K. The timestep was set as 0.5 fs. In the simulation, the NVT ensemble was fixed first for 100,000 steps and then followed by an NVE ensemble for 100,000 steps to fully relax the lattice. Then the system was fixed at the NVE ensemble for another 200,000 steps, during which the temperature gradient and heat flow rate were recorded. Then the thermal conductivity can be obtained based on Fourier's law. Figure S6b shows an example of the temperature gradient along the *x*-axis for the $\delta = 0.25$ case. It is noticed that the hot reservoir has the highest temperature, while the cold reservoir exhibits the lowest temperature. The length between the hot and cold reservoir is 9.2 nm, different from that in Fig. S6a. This discrepancy is caused by the position where the temperature is monitored not being precisely at the boundary of the reservoirs. Figure S6c displaces the cumulative heat flow applied to or extracted from the system. Linear fitting is applied to obtain the heat flow rate. The heat flow rate and temperature difference on both sides are averaged when evaluating the thermal conductivity.

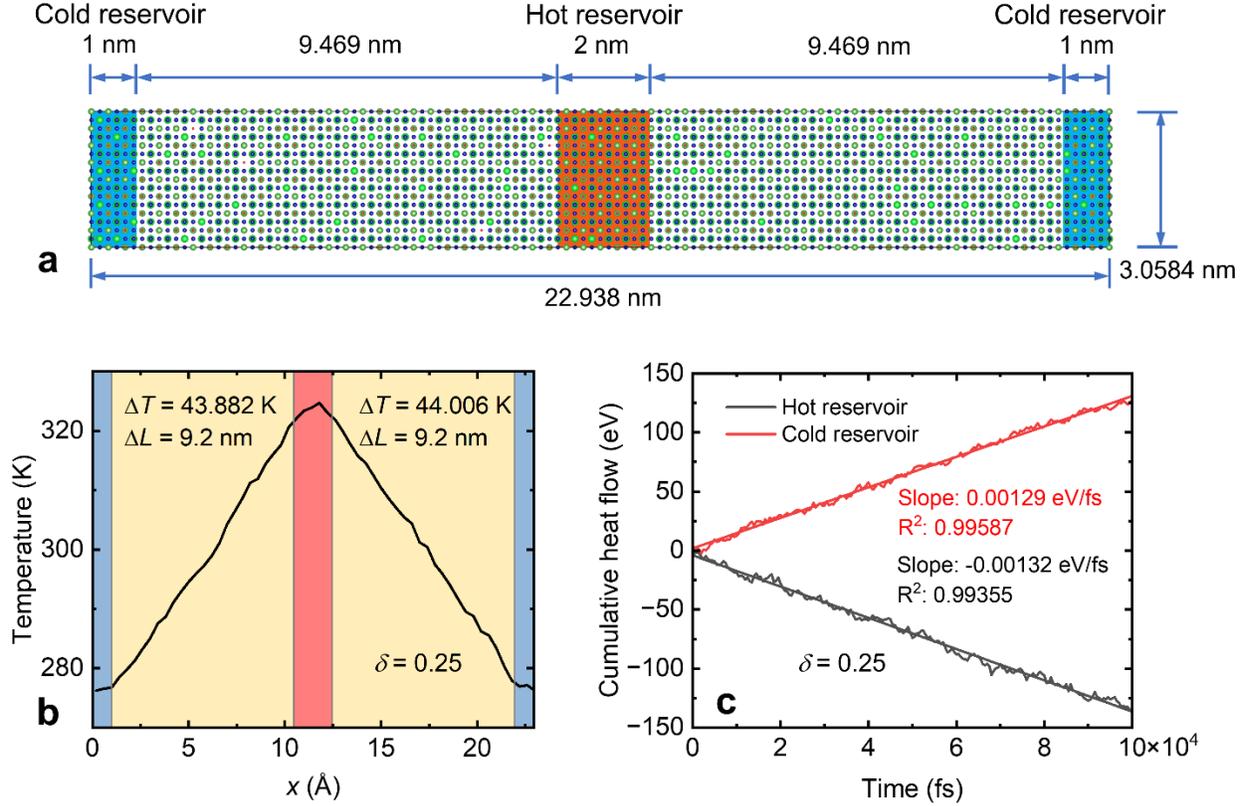

**Fig. S6 | NEMD setup and results**. **a** Dimensions of the NEMD simulation system, and the hot and cold reservoirs in the *x-y* plane. **b** Temperature gradient along the *x* direction for the $\delta = 0.25$ case. **c** Cumulative thermal energy in the hot and cold reservoirs.

## 6. Temperature-dependent thermal measurements

For data analysis of temperature-dependent thermal measurements, we need to obtain the temperature-dependent volumetric heat capacities of the Pt transducer, the LSCO film, and the LAO substrate ($C_{Pt}$, $C_{LSCO}$, $C_{LAO}$), and the temperature-dependent thermal conductivities of Pt and LAO ($\Lambda_{Pt}$ and $\Lambda_{LAO}$) as input parameters. $C_{Pt}$ and $\Lambda_{Pt}$ were obtained from the literature[17-19]. $C_{LAO}$ and $C_{LSCO}$ were calculated using the Debye model following:

$$C = 5 \times \frac{9\bar{R}\rho}{M} \left(\frac{T}{\theta_D}\right)^3 \int_0^{\frac{\theta_D}{T}} \frac{x^4 e^x dx}{(e^x - 1)^2}, \qquad (4)$$

where $R$ is the universal gas constant (8.314 J mol$^{-1}$ K$^{-1}$), $\rho$ is the mass density, $M$ is the molar mass, $T$ is temperature, and $\theta_D$ is the Debye temperature (748 K for LAO[19] and 462 K for LSCO[20]).

The calculated values of temperature-dependent $C_{LAO}$ and $C_{LSCO}$ are presented in Fig. S7a in comparison with literature data[19,21-23].

We measured a bare LAO substrate with the same Pt transducer to obtain the temperature-dependent thermal conductivity of LAO. For thermal conductivity measurements at low $T$, temperature correction is necessary considering that material properties may vary significantly over a small temperature rise. The temperature correction takes into account two sources of temperature rise: (1) the steady-state temperature rise ($\Delta T_s$), and (2) the per-pulse pump heating induced temperature rise ($\Delta T_p$). The calculation of $\Delta T_s$ can be found in Ref. [24]. $\Delta T_p$ was calculated as $\Delta T_p = Q/(VC_{Pt})$, where $Q$ is the pump energy absorbed by the Pt transducer ($Q = 2a_{Pt}A_{Pump}/f_{laser}$, with $a_{Pt}$ being the absorptivity of Pt, $A_{pump}$ being the absorbed pump power, and $f_{laser}$ being the repetition rate of the laser), $V$ is the volume of a cylinder (with a $1/e^2$ radius of the pump beam spot size and height of the Pt thickness), and $C_{Pt}$ is the volumetric heat capacity of Pt. The corrected temperature was calculated as $T_{corr} = T_{set} + \Delta T_p + \Delta T_s$ with $T_{set}$ being the setting temperature set in the temperature controller. Those input parameters were then modified based on $T_{corr}$ for a new round of data reduction. We iterated this correction process several times until we obtained converged results (*e.g.*, $\Delta T_p + \Delta T_s \leq 15$ K).

Figure S7b shows the comparison between our measured $\Lambda_{LAO}$ and literature values of $\Lambda_{LAO}$ along different crystalline orientations from experimental studies[19,21-23,25]. Due to the existence of twin boundaries in LAO, the thermal conductivity of LAO along the [100] direction is much lower than that along the [001] direction (black symbols) and the temperature-dependent thermal conductivity deviates from the $1/T$ trend. Our measured temperature-dependent $\Lambda_{LAO}$ along the [100] direction agrees reasonably well with literature data. The measured $\Lambda_{LAO}$ values were then used as input parameters for LSCO measurements at different temperatures.

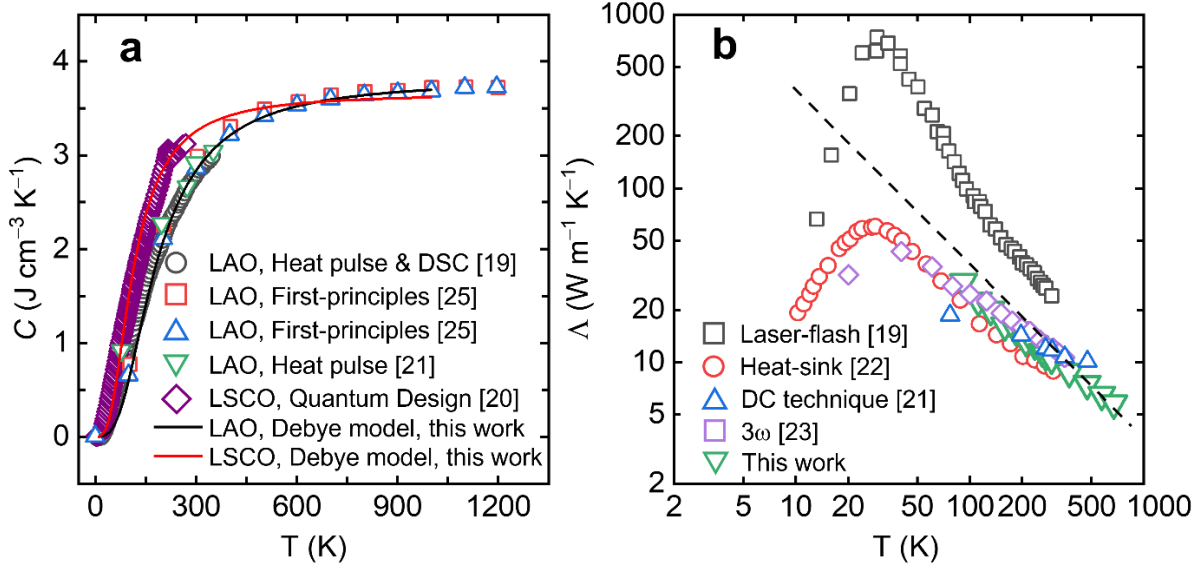

**Fig. S7 | *T*-dependent heat capacity and thermal conductivity. a** *T*-dependent heat capacity of LAO[19,21-23] and LSCO[20]. Note that in Ref. [20], $x = 0.3$ is the highest $x$ studied for single-crystal La$_{1-x}$Sr$_x$CoO$_3$, and is thus the closest sample system to our La$_{1-x}$Sr$_x$CoO$_3$ ($x = 0.5$) film available in the literature. **b** *T*-dependent thermal conductivities of LAO obtained from different measurement methods[19,21-23,25]. The black dashed line indicates the $1/T$ trend of temperature-dependent thermal conductivity.

In addition to the BM-LSCO film gated to 3 V, we also conducted temperature-dependent thermal measurements of the BM-LSCO film gated at 4 V and the results are plotted in Fig. S8. The thermal conductivity of the sample gated to 4 V (red diamonds) agrees well with that of the sample gated to 3 V (blue circles) over the measurement temperature range of ~ 90 to 500 K. This further supports, from the thermal characterization aspect, our finding that LSCO is fully transformed to the BM phase after being gated at 3 V and above.

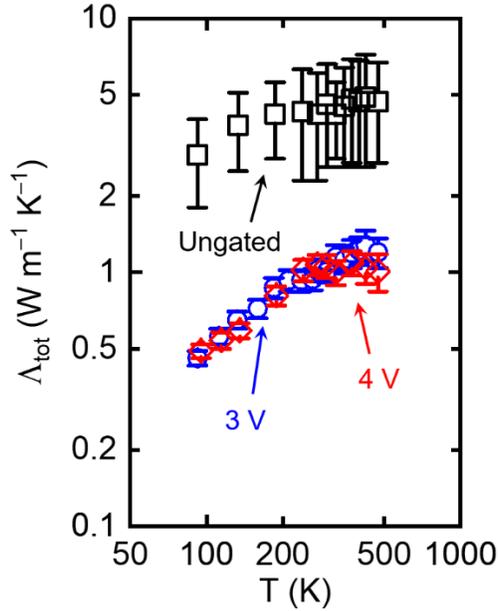

**Fig. S8 | *T*-dependent thermal conductivity.** Temperature-dependent total thermal conductivities of LSCO films at different gate voltages.

## 7. Comparison of in-plane vs. through-plane transport properties

Due to the low measurement sensitivity to the in-plane thermal conductivity of LSCO thin films at these types of thicknesses[26], in addition to the challenges of probing the through-plane electrical resistivity, we were unable to measure the electronic and thermal transport along the same direction. However, considering that the thickness of our measured samples is ~45 nm, we can safely conclude that the electronic transport in our P-LSCO samples exhibits bulk behavior. Estimates of the electronic mean free path in this system yield nanometric scales. Also, from *ab initio* calculations of P-SrCoO$_3$ in a prior study[27], the accumulated lattice thermal conductivity of P-SrCoO$_3$ is larger than 93% of its bulk value when the phonon mean free path reaches 40 nm. In our P-LSCO system, with the additional alloying on the A site, the lattice thermal conductivity will approach the bulk limit at an even smaller film thickness. In addition, considering the cubic symmetry of P-LSCO, we expect quite isotropic band structures for both electrons and phonons.

Thus, the through-plane thermal conductivity of P-LSCO, consisting of both electronic and lattice contributions, must be comparable to the in-plane thermal conductivity.

Unlike P-LSCO, BM-LSCO is electrically insulating; therefore, thermal transport in BM-LSCO is mainly carried by phonons and will also approach the bulk limit. However, there is possible anisotropy for thermal transport in BM-LSCO due to the orthorhombic symmetry of the BM phase. In this case, the large tuning factor of the LSCO thermal conductivity we demonstrated in this work is along the through-plane ($c$-axis) direction.

## 8. Reversibility of tuning the LSCO thermal conductivity

To explore the reversible nature of the $V_g$-induced P $\leftrightarrow$ BM phase transformation in LSCO, we also reverted some gated BM-LSCO films to the P phase by applying reverse gate voltages of up to $-4.5$ V at RT. The RT thermal conductivity of one reverse-gated P-LSCO film (P $\rightarrow$ BM $\rightarrow$ P) was measured to be $3.5 \pm 1.1$ W m$^{-1}$ K$^{-1}$. This ~20% reduction in the $\Lambda_{tot}$ of recovered P-LSCO compared to the as-deposited P-LSCO film arises from additional structural disorder induced during the cyclic gating. This can be seen from the high-resolution XRD measurements (Fig. S9a,b) on the as-deposited P-LSCO film (labeled as "ungated"), after gating to $+3.0$ V (*i.e.*, transformed to BM), and after reverse gating to $-4.5$ V (*i.e.*, recovered to P). The ideal situation after negative $V_g$ would be a P(002) reflection exactly overlapping with the ungated peak. In reality, the return to the ungated P(002) location from the BM(008) position has a remnant shift to the left in Fig. S9b (*i.e.*, a larger P lattice parameter). Explicitly, the BM(006) and (0010) peaks are entirely extinguished, definitively establishing that long-range $V_O$ order is annihilated, but the lattice parameter of the recovered P phase remains expanded. This indicates that the $V_O$ concentration is higher than it was originally for the ungated sample. This additional structural disorder can be further quantified through electronic transport (Fig. S9c). Specifically, the approach described in the main text and Methods of our paper to determine the O non-stoichiometry $\delta$ from transport yields 0.11 for the as-deposited P-LSCO film and 0.16 for the reverse-gated film (following P → BM → P). This additional disorder is unsurprising after topotactic transformation from P to BM back to P, and no doubt plays some role in the reduced thermal conductivity. Nevertheless, such data confirm the overall reversibility of the approach presented here. Further improvement of reversibility can be achieved through better device design and optimized gating conditions.

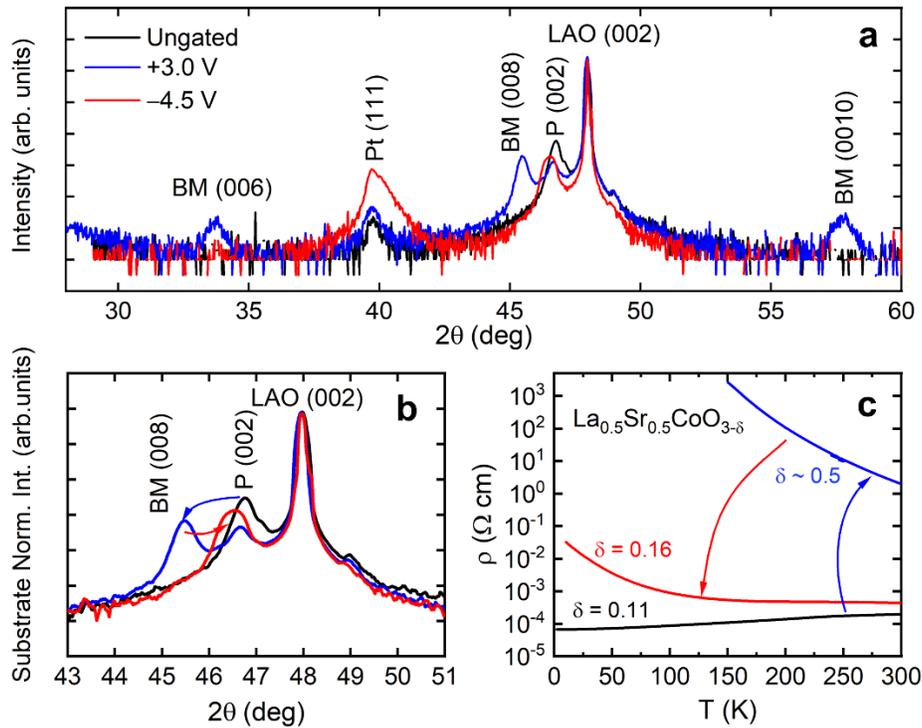

**Fig. S9 | XRD measurements on a reverse-gated sample. a** Wide-angular-range XRD results from LSCO devices on LAO(001) substrates after gating to +3.0 V and after reverse gating to −4.5 V. The *y*-axis is on a logarithmic scale. **b** An enlarged version of the same scans as in a, around the film and substrate perovskite (002) peak (BM (008)), with the intensity axis (log scale) normalized to the substrate intensity, to better show the changes in the film position and intensity with reverse gating. **c** Temperature-dependent electrical resistivity measurements of the as-deposited P-LSCO, after gating to +3.0 V, and after reverse gating to −4.5 V (the color code is the same for all panels).